\documentclass[pre,aps,
twocolumn,pdflatex,
superscriptaddress,floats,showpacs,10pt]{revtex4-1}
\usepackage{graphicx}
\usepackage{dcolumn}
\usepackage{bm}
\usepackage{amssymb}
\usepackage{amsmath}
\usepackage{array}
\usepackage{float}
\usepackage{geometry}
\usepackage{color}
\usepackage{amsthm}

\begin{document}

\title{Prepulse and amplified spontaneous emission effects on the
interaction of a petawatt class laser with thin solid targets}
\author{ Timur Zh. Esirkepov$^{a}$, James K. Koga$^{a}$, Atsushi Sunahara$^{b}$, Toshimasa Morita$^{a}$,\\
Masaharu Nishikino$^{a}$, Kei Kageyama$^{b}$, Hideo Nagatomo$^{c}$, Katsunobu Nishihara$^{c}$, Akito Sagisaka$^{c}$,\\
Hideyuki Kotaki$^{a}$, Tatsufumi Nakamura$^{a}$, Yuji Fukuda$^{c}$, Hajime Okada$^{a}$, Alexander Pirozhkov$^{a}$,\\
Akifumi Yogo$^{a}$, Mamiko Nishiuchi$^{a}$, Hiromitsu Kiriyama$^{a}$, Kiminori Kondo$^{a}$,\\
Masaki Kando $^{a}$, Sergei V. Bulanov$^{a,d,e}$\\
{\small $^{a}$Kansai Photon Science Institute, JAEA, Kizugawa, Kyoto 619-0215, Japan}\\
{\small $^{b}$Institute for Laser Technology, 2-6 Yamadaoka Suita Osaka 565-0871, Japan}\\
{\small $^{c}$Institute of Laser Engineering, 2-6 Yamadaoka Suita Osaka 565-0871, Japan}\\
{\small $^{d}$Prokhorov Institute of General Physics, RAS, Moscow 119991, Russia}\\
{\small $^{e}$Moscow Institute of Physics and Technology, Dolgoprudny, }\\
{\small Moscow region 141700, Russia}
}


\date{\today }

\begin{abstract}
When a finite contrast petawatt laser pulse 
irradiates a micron-thick foil,
a prepulse (including amplified spontaneous emission)
creates a preplasma,
where
an ultrashort relativistically strong portion of the laser pulse 
(the main pulse)
acquires higher intensity due to relativistic self-focusing
and undergoes fast depletion transferring energy to fast electrons.  
If the preplasma thickness is optimal,
the main pulse can reach the target
generating fast ions 
more efficiently than an ideal, infinite contrast, laser pulse.
A simple analytical model of a target with preplasma formation is developed
and the radiation pressure dominant acceleration of ions
in this target is predicted.
The preplasma formation by a nanosecond prepulse
is analyzed with dissipative hydrodynamic simulations.
The main pulse interaction with the preplasma is
studied with multi-parametric particle-in-cell simulations. 
The optimal conditions for hundreds of MeV ion acceleration
are found
with accompanying effects important for diagnostics, 
including high-order harmonics generation.
\end{abstract}


\maketitle

\section{Introduction}
\label{sec:Introduction}

Petawatt (PW) power class laser interaction with various targets 
enables novel regimes of high energy charged particle acceleration
and high brightness coherent and incoherent electromagnetic radiation
generation over a wide range of photon energies 
\cite{MTB, KI, ESL, SC, UFN, HHG}.

One of the central scientific goals of studying relativistic laser plasmas
is to obtain high-quality ion beams accelerated to hundreds mega-electron-volt (MeV) per nucleon,
because this is a crucial milestone on the road towards 
the laser ion accelerator for applications in hadron therapy \cite{BK}. 
Various laser ion acceleration mechanisms have
been discussed in theoretical and experimental papers
(see review articles \cite{MTB, BFB, DNP, MBP} and literature cited therein).

Apparently the maximum ion energy increases with the laser focused intensity
which is in turn proportional to the laser power.
Protons of 60 MeV from
100 $\mu$m foils irradiated by 400 J subpicosecond laser pulses of a PW class
laser system have been detected in Ref. \cite{Snavely}. 
With laser pulses under 10 J,
the highest proton energy obtained so far is 40 MeV  \cite{Ogura},
for a micron-thick metal foil
irradiated by an ultra-short 200 TW femtosecond pulse laser 
($\tau_{las}\approx 40$fs) at an intensity of about $10^{21}$W/cm$^2$.
A 85 MeV proton generation has been detected with the petawatt laser
at APRI-GIST, Korea \cite{Jeong}.

According to the theoretical concept formulated in Ref. \cite{TARA},
a femtosecond petawatt class laser pulse
focused onto a thin solid density proton-containing target
can blow off almost all the electrons
creating a Coulomb potential which accelerates protons
to the energy of $\mathcal{E}_p$,
which scales with the laser power, $\mathcal{P}$, as
\begin{equation} \label{eq:tara}
\mathcal{E}_p=m_e c^2 
\sqrt{\chi \mathcal{P}/\mathcal{P}_{rel}} \approx
173 \sqrt{\chi \mathcal{P} [\mbox{petawatt}]}\, \mathrm{MeV}.  
\end{equation}
Here the coefficient $\chi$ is of the order of unity
and depends on the laser pulse shape and its energy absorption;
$\mathcal{P}_{rel}= m_e^2c^5/e^2\approx 8.71\,$GW is proportional to 
the critical power for relativistic self-focusing \cite{GZSUN};
$e$ and $m_e$ are the electron charge and mass;
$c$ is the speed of light in vacuum.
The laser radiation with the power of $\mathcal{P}_{rel}$ 
focused into a spot with the diameter of the laser wavelength, $\lambda$,
produces the intensity of $I_{rel}\approx 0.87\times 10^{18}\,$W/cm$^2\times(1\,\mu$m$/\lambda)^2$
and the corresponding electric field of $E_{rel} = 2 m_e\omega c/\pi e$
reaching the relativistic limit \cite{MTB}.
In terms of the dimensionless amplitude, 
$a=eE/m_e\omega c = 0.85\sqrt{I_{las}[\mbox{exawatt/cm}^2]}(\lambda [\mu\mbox{m}])$,
where $\omega=2\pi c/\lambda$ is the laser frequency and $I_{las}$ is the focused intensity,
the relativistic limit is $a=1$.
The relationship in Eq. (\ref{eq:tara}) shows that the 200 MeV
proton energy can be achieved with $\approx 1.3\,$PW laser power on the target.

A femtosecond petawatt laser pulse (the {\it main pulse})
obtained with the present-day laser technology \cite{MTB}
is typically accompanied by a 
relatively low-energy nanosecond {\it prepulse}
which is a combination of 
sub-picosecond pulses, a picosecond ramp and
nanosecond amplified spontaneous emission (ASE).
The {\it prepulse} heats, melts and evaporates 
a portion of an initially solid density target
creating a preplasma at the target front
on the timescale of nanoseconds.
The {\it main pulse} then interacts with the preplasma
before it can reach the solid density region. 
These effects can substantially modify the 
laser -- thin solid target interaction
(e.g., see experimental and theoretical results on the ion acceleration 
in Refs. \cite{MATSU, YOGO},
where the {\it prepulse} transforms a micron foil into 
a finite thickness near-critical plasma layer,
and in Ref. \cite{WANG}, 
where a low-contrast 
of a 3 TW  
{\it main pulse} impedes ion acceleration).
Research into optimization of the prepulse for picosecond laser-irradiation
of thin foil targets has been previously performed \cite{MCKENNA}.

A fast depletion of the {\it main pulse} propagating in preplasma 
can diminish the ion acceleration efficiency.
However the {\it main pulse} also undergoes relativistic self-focusing
which increases its intensity and 
decreases the volume of the laser field immediate interaction (tightening the pulse waist).
A mutual counteraction
of these effects can lead to the enhancement of ion acceleration efficiency,
as we show below.

In this paper we investigate how the {\it prepulse} modifies
the interaction of petawatt class laser with thin solid targets.
We find optimal conditions for the ion acceleration
and reveal accompanying effects which are useful for diagnostics in
experimental searches for the optimal regimes.

In order to accomplish this task, we
performed two kinds of numerical simulations strongly separated by the timescale.
The study of the interaction of a nanosecond sub-mJ {\it prepulse} with a micron foil
has required simulations using dissipative hydrodynamic algorithms
described in Refs. \cite{UTSUMI, NAGATOMO, SUNAHARA,OHNISHI}.
The interaction of a femtosecond several joule {\it main pulse}
with preplasma modelled following dissipative hydrodynamics simulations
have been studied with multi-parametric particle-in-cell (PIC) simulations
similarly to Refs. \cite{MATSU,TZE}.

In the present paper we 
briefly review the ion acceleration mechanisms in Sec. \ref{sec:IAM};
describe typical parameters of the laser pulse components in Sec. \ref{sec:ASE};
formulate a simple analytical model for preplasma formation and ion acceleration in Sec. \ref{sec:Model};
present dissipative hydrodynamic simulation results in Sec. \ref{sec:HD};
summarize the multi-parametric PIC simulation results in Sec. \ref{sec:PIC};
discuss the outcomes and conclude in Sec. \ref{sec:Conclusion}.


\section{Ion Acceleration Mechanisms}
\label{sec:IAM}

Several basic laser ion acceleration mechanisms have been established,
depending on the laser pulse and target parameters.
They can be categorized into several groups.

The most actively studied regime so far is Target Normal Sheath
Acceleration (TNSA) \cite{TNSA}. TNSA is realized, when a low contrast
laser pulse interacts with a thick solid density slab target. This implies
that a portion of the laser-heated electrons leave the target in the forward
direction with respect to the laser pulse propagation and establish a
longitudinal electric field at the target rear surface. In the quasistatic
limit the ion energy is determined by the electrostatic potential there. In
the dynamical regime not only hot electrons leave the target but also a
plasma cloud formed at the target rear side expands into vacuum. The
fast ions are accelerated at the front of the expanding plasma cloud
\cite{Gurevich-P-Pitaevsky}. The resulting ion beam has a broad quasi-thermal energy
spectrum with a cut-off. Most of the experimental results on laser proton acceleration
obtained so far, including the above mentioned 40 MeV proton acceleration seen in Ref. \cite{Ogura},
can be attributed to the TNSA scheme.

A high contrast strong enough laser pulse can push away all the electrons
from a thin or mass limited solid density target 
in a time shorter than the ion response time.
Then ions undergo a Coulomb explosion  due to the repulsion of the
positive electric charge \cite{LAST}. The resulting ion beam has
a nonthermal energy distribution with a cutoff at the energy 
determined by the maximum of the electrostatic potential of the ion core.
The ion energy scaling given by Eq. (\ref{eq:tara}) corresponds to
this regime, and can be realized
when the laser pulse irradiates a thin double layer target
(consisting of high-Z layer and a much thinner proton-containing layer).
In the Coulomb explosion of high-Z layer, protons acquire the highest energy.
The double layer target can secure obtaining high-quality
(quasi-monoenergetic and low-emittance) ion beams, \cite{BK,TARA,DoubleLayer,TZE}.
The proton-containing layer can be prepared in a
controllable way as in Ref. \cite{SCHW} or
can be a water contamination layer usually present on metal foils.

Radiation Pressure Dominated Acceleration \cite{RPDA}, comes into play when
the laser is able to push the foil as a whole by the electromagnetic radiation
pressure. This mechanism is a realization of the relativistic receding
mirror concept \cite{UFN}. The laser pulse is reflected by a co-moving
mirror with its energy transferred to the mirror. Recently, several papers
have reported on the experimental indication of the onset of this regime of
laser ion acceleration, \cite{KAR}. The transition from the TNSA to
the RPDA regime has been observed in the PW class laser beam interaction 
with nanoscale solid foil targets, when 45 MeV
protons have been detected \cite{IJKim}.

The Magnetic Vortex Acceleration \cite{MVA} regime occurs when the laser
interacts with a near-critical density target, where it makes a channel in
both electron and ion density. Exiting from the plasma the laser pulse
establishes a strong longitudinal electric field sustained by a quasistatic
magnetic field associated with the vortex motion of electrons. 
This electric field accelerates the ions. In the case of
sub-picosecond pulses the acceleration of helium ions up to 40 MeV from underdense
plasma by the VULCAN laser \cite{WIL1} and the acceleration of protons up
to 50 MeV by the Omega EP laser \cite{WIL2} have been observed.
Also experiments with femtosecond pulses irradiating cluster jet targets
show that 10-20 MeV per nucleon ions can be generated \cite{Fukuda}.

A combination of the basic laser ion acceleration mechanisms
can enhance the maximum ion energy, or increase the number of accelerated ions,
or modify the ion beam spectrum. For example, the
accelerated ion energy can be substantially increased with the Directed
Coulomb Explosion scheme \cite{Stepan-DC}, which is the
combination of Radiation Pressure Acceleration and Coulomb Explosion
mechanisms.  Another example can be found in \cite{MORITA}.

Target micro- (and nano-) structuring can enhance the laser pulse coupling
with the target \cite{DM,SK} and improve the accelerated ion beam quality
\cite{BK, TARA, DoubleLayer,SCHW}.

In the case when the laser radiation interacts with an undercritical
density target the ponderomotive pressure of a wide enough laser pulse
launches a collisionless shock wave propagating in the forward direction.
The ions reflected at the shock wave front acquire a velocity twice the
velocity of the shock, producing a narrow energy spectrum ion beam
\cite{SHOCK}. The accelerated ion beam interacting with the background
plasma becomes subject to two-stream and filamentation instabilities
\cite{SLOW}, the latter determines the width of the accelerated ion
spectra.


\section{Parameters of the Main Pulse, Prepulse and Amplified Spontaneous Emission}
\label{sec:ASE}

In the analysis of the laser -- thin solid target interaction
we consider laser pulses of the petawatt class.
As can be inferred from experimental and theoretical results mentioned in previous sections,
such the power and the focused intensity of $10^{21}\,$W/cm$^2\times(1\,\mu{\rm m}/\lambda)^2$
are required for the laser-driven generation of hundreds of MeV ions.
We consider the laser pulse energy, $\mathcal{E}_{las}$, of the order of tens of joules,
which entails the {\it main pulse} duration, $\tau_{las}$, of the order of tens of femtoseconds.

\begin{figure}[tbh]
\includegraphics[width=0.99\columnwidth]{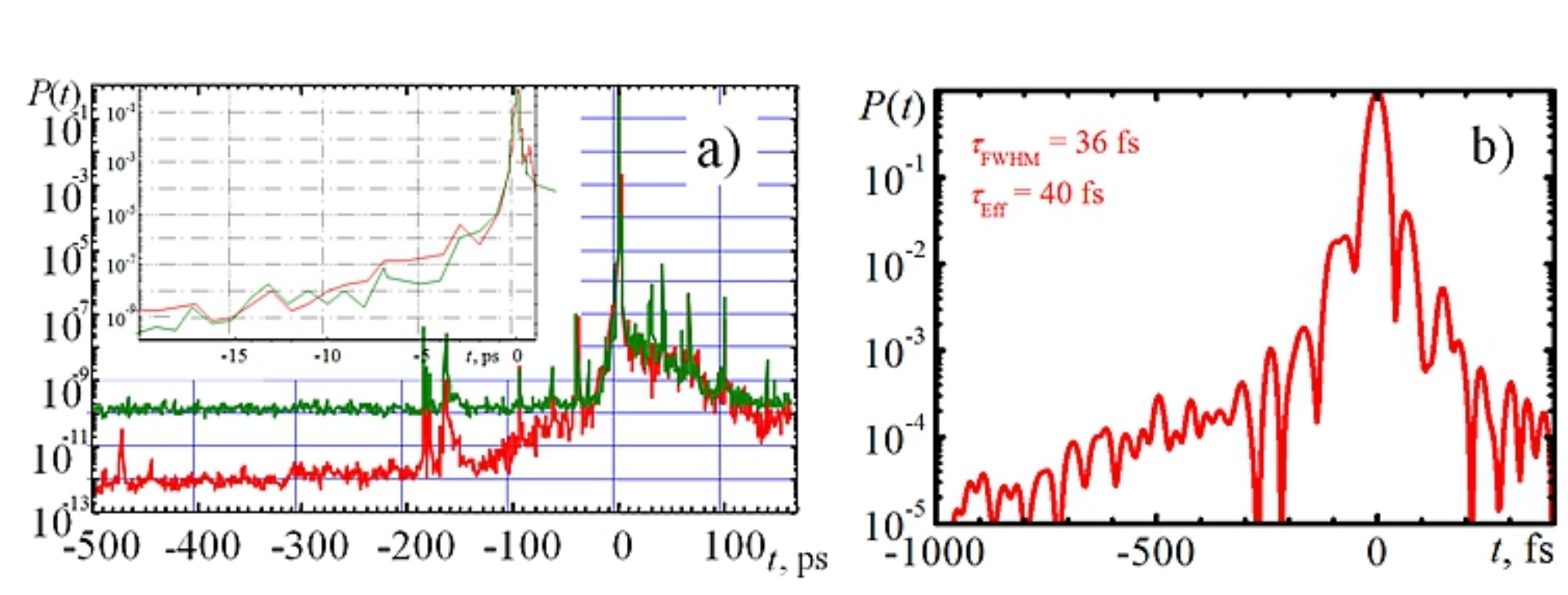} 
\caption{\label{fig:J-KAREN}
Time dependence of the J-KAREN laser pulse power in the 
a) nanosecond and b) picosecond range, \cite{KIRIYAMA2012}.
}
\end{figure}

As an example of a typical structure of a petawatt class laser pulse,
in Fig. \ref{fig:J-KAREN} we show the time dependence of the pulse power for 
the J-KAREN laser \cite{KIRIYAMA2012} used in Ref. \cite{Ogura}.
The {\it main pulse} containing the main part of the energy 
has the duration of $\tau_{las}\approx 40\,$fs.
It is preceeded by radiation with much less power, which we call the {\it prepulse},
with the duration of the order of $\tau_{pp}\approx 1\,$ns
and total energy, $\mathcal{E}_{pp}$, less than mJ.
The main components of this long timescale radiation
are the nanosecond amplified spontaneous emission (ASE) 
and a few picosecond ramp.
The peaks in Fig. \ref{fig:J-KAREN} a few tens and hundreds of ps before the main pulse originate from post-pulse transfer to prepulse due to self-phase modulation at the chirped stage of the CPA amplifier chain \cite{DIDENKO}. Some of these peaks can also be artifacts generated in the cross-correlator.
The few picosecond ramp originates from the high-order phase modulations
determined by high-order dispersion terms.

The laser pulse contrast in terms of energy
is characterized by the ratio of the {\it main pulse} to the {\it prepulse} energy,
\begin{equation}
C_{\mathcal{E}}=\mathcal{E}_{las}/\mathcal{E}_{pp},
\label{eq:CE}
\end{equation}
which we consider in the range from $10^4$ to $10^6$.
We introduce also the intensity contrast equal to the ratio of the {\it main pulse}
to the {\it prepulse} intensity, 
\begin{equation}
C_{I}={I}_{las}/{I}_{pp}.
\label{eq:CI}
\end{equation}
When the {\it main pulse} intensity is $I_{las} = 10^{21}\,$W/cm$^2$,
the intensity contrast is $10^{9}$ for the {\it prepulse} intensity of $I_{pp}=10^{12}$W/cm$^2$, 
and is $10^{11}$ for $I_{pp}=10^{10}$W/cm$^2$.

We note that the preplasma formation has been studied
in Ref. \cite{SAGISAKA} experimentally and via hydrodynamics simulations
for a few orders of magnitude lower laser energy.


\section{Simple Model for Preplasma Formation and Ion Acceleration}
\label{sec:Model}

\subsection{Simple model for preplasma formation}

When laser radiation interacts with a solid target, the {\it prepulse} produces
an extended preplasma. The key parameters of the preplasma are
the electron temperature, plasma density and the expansion velocity, and the
preplasma spatial scale.

The electron temperature, $T_e$, grows due to collisional heating of the
plasma undergoing irradiation by the laser {\it prepulse} field, $E_{pp}$, 
oscillating with frequency $\omega$. 
The minimal electron temperature can be estimated to be equal to
the quiver electron energy: 
\begin{equation}
\min \{{T_e}\}=\min \{\mathcal{E}_e\} =m_e c^2\sqrt{
{a_0^2}/{C_I}},
\label{eq:Tmin}
\end{equation}
where $C_I = E_0^2/E_{pp}^2$ is the {\it prepulse} intensity contrast 
determined by Eq. (\ref{eq:CI}), and $a_0 = eE_0/m_e\omega c$
is the dimensionless amplitude of the {\it main pulse}.

In a collisional plasma the electron temperature growth is determined by
equation \cite{RAIZER} 
\begin{equation}\label{eq:DRU}
\frac{d T_e}{dt}=\left(\frac{e^2 E_{pp}^2}{m_e (\omega^2+\nu^2)}-\frac{2 m_e}{m_i}T_e \right)\nu,
\end{equation}
with the collision frequency, $\nu$, which depends on the electron temperature as 
$\nu (T_e)=4.4\times 10^{12} \mbox{s}^{-1} \times (n_e/n_e^*) (T_e/T_e^*)^{-3/2}$.
For an electron density and temperature of the order of 
$n_e^*=10^{21}$cm$^{-3}$ and $T_e^*=100\,$eV,
the collision frequency is $\nu=4.4\times 10^{12}$s$^{-1}$.
Table \ref{TAB2}  shows the collision frequency value 
for different values of the electron density and temperature.

\begin{table}[h]
\begin{tabular}{|c|c|c|c|}
\hline
$\nu$ [s$^{-1}$] & $T_e=100$           & $T_e=250$           & $T_e=1500$       \\ \hline
$n_e=10^{20}$    & $4.4\times 10^{11}$ & $1.1\times 10^{11}$ & $7.6\times 10^{9}$  \\ \hline
$n_e=10^{21}$    & $4.4\times 10^{12}$ & $1.1\times 10^{12}$ & $7.6\times 10^{10}$ \\ \hline
$n_e=10^{22}$    & $4.4\times 10^{13}$ & $1.1\times 10^{13}$ & $7.6\times 10^{11}$ \\ \hline
\end{tabular}
\caption{\label{TAB2}
Collision frequency, $\nu$, for different values of the 
electron density (in cm$^{-3}$) and temperature (in eV).}
\end{table}

In the long time limit, $\nu t\gg m_{i}/2m_{e}$, Eq. (\ref{eq:DRU}) yields 
\begin{equation}
T_{e} =
(m_{i}/2m_e)(e^{2}E_{pp}^{2} / m_{e}\omega^2) =
m_i c^2 a_0^{2} / 2C_{I} .
\label{eq:TeLong}
\end{equation}
For $0\ll \nu \Delta t \ll m_{i}/2m_{e}$,
from Eq. (\ref{eq:DRU}) we have 
$T_{e} \approx 
\left[ 1.1\times 10^{16} (n_e/n_e^*) (e^2 E_{pp}^2 / m_e\omega^2) \Delta t \right]^{2/5}$,
where the time period $\Delta t$ is in seconds 
and the unit of $m_e$ is eV/$c^2$;
\begin{equation} \label{eq:TeSHRT}
T_{e} \approx 
1.26\times 10^{5}
\left( \frac{n_e}{10^{21}\mbox{cm}^{-3}} \frac{\Delta t}{1\mbox{ns}} \frac{a_0^2}{C_{I}} \right)^{2/5} \mbox{eV}.
\end{equation}

As seen from Table \ref{TAB2},
presenting the collision frequency in a wide range of the electron density and temperature,
for a nanosecond {\it prepulse} we are in the regime of $0\ll \nu \Delta t \ll m_i/2 m_e$.

Using Eqs. (\ref{eq:Tmin}) 
and (\ref{eq:TeSHRT}) 
we obtain an interpolation
\begin{equation}
T_{e}=
\left[
5.1\sqrt{\frac{a_0^{2}}{C_{I}}}
+
1.3\left( \frac{n_e}{10^{21}\mbox{cm}^{-3}} \frac{\Delta t}{1\mbox{ns}} \frac{a_0^2}{C_I} \right)^{2/5}
\right]\times 10^{5}\mbox{ eV}
\, .
\label{eq:TeINTRPL}
\end{equation}

In Table \ref{TAB1}  we present the electron temperature for different
values of the {\it main pulse} amplitude and contrast $C_{I}$,
for a nanosecond {\it prepulse} and the electron density of $n_{e}=10^{21}$cm$^{-3}$.

\begin{table}[h]
\begin{tabular}{|c|c|c|c|}
\hline
$T_e$ [eV] & $C_I=10^9$ & $C_I=10^{10}$ & $C_I=10^{11}$ \\ \hline
$a_0=$10   & 370        & 130           & 50            \\ \hline
$a_0=$40   & 1270       & 450           & 160           \\ \hline
$a_0=$100  & 3000       & 1000          & 370           \\ \hline
\end{tabular}
\caption{\label{TAB1}
Electron temperature, $T_e$, versus
the {\it main pulse} amplitude $a_0$ and contrast $C_{I}$,
according to Eq. (\ref{eq:TeINTRPL}). 
The electron density is $n_e=10^{21}$ cm$^{-3}$, 
the {\it prepulse} duration is $\tau_{pp} = \Delta t = 1$ ns.
}
\end{table}

\begin{figure}[bh]
\includegraphics[width=0.8\columnwidth]{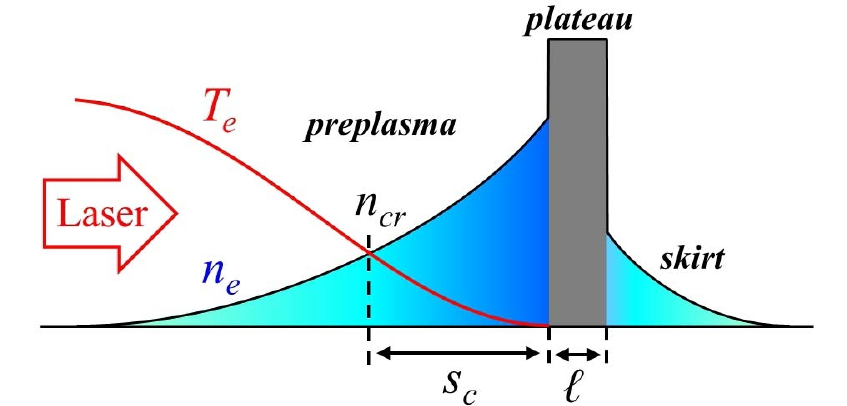}
\caption{\label{fig:prep-plat-skirt}
Schematic distributions of the electron density and temperature
in a typical target structure created by the nanosecond {\it prepulse}.
The remaining portion of the foil, {\it plateau},
is surrounded by a {\it preplasma} at the front side 
and by a {\it skirt} at the back side.
}
\end{figure}

Fig. \ref{fig:prep-plat-skirt} 
shows a schematic distribution of the electron density 
and temperature in the target modified by the {\it prepulse}.
The remaining portion of the foil
with a high density, {\it plateau},
is surrounded by a relatively low density 
{\it preplasma} at the front and a {\it skirt} at the rear.

The {\it prepulse} energy is absorbed mainly in the region close to the critical density surface,
$r=s_c$, where $n_e(s_c)=n_{cr}=m_e \omega^2/4 \pi e^2$.
This causes the electron heating with the 
heat transport due thermal conductivity towards the overcritical density region, 
the ion heating and the preplasma expansion. 
A part of the absorbed energy is radiated 
away due to radiation losses. 
One can find a similarity between this scenario and physical processes 
which occur during nanosecond laser radiation 
interaction with solid targets \cite{Bodner, Lindl}. 

Now we estimate the ``cost'' of the preplasma formation, ${\cal E}_{C}$, 
considering the energy balance with the prepulse energy ${\cal E}_{pp}$.
We write it  
in the form of a sum
\begin{equation}
{\cal E}_{C}=N_e T_e+N_i T_i+N_e {\cal E}_I+N_i {\cal E}_{i,{\rm kin}}+{\cal E}_{\rm rad}
\end{equation}
where $N_e$ and $N_i$ are the total number of electrons and ions in the preplasma; 
$T_e$ and $T_i$ are the electron and ion temperature;
${\cal E}_I$ is the ionization ``cost'' approximately equal to 
twice the ionization potential;
${\cal E}_{i,{\rm kin}}$ is the ion kinetic energy, 
and ${\cal E}_{\rm rad}$ is the energy radiated away. 
The energy ${\cal E}_{C}$  
is equal to the absorbed prepulse energy
\begin{equation}
{\cal E}_{C}=\kappa_{\rm abs} {\cal E}_{pp},
\end{equation}
where $\kappa_{\rm abs}$ is the absorption coefficient. 

Assuming the absorption coefficient to be of the order of 0.3 and 
the {\it prepulse} energy of ${\cal E}_{pp}\approx 1\mbox{ mJ} = 6.24\times 10^{15}\,$eV 
we find that for the electron temperature of the order of 300 eV,
the total electron number in the preplasma is $N_e\approx 2\times 10^{13}$. 
Since these electrons originate from the foil with the solid density, 
$\approx 10^{24}$cm$^{-3}$, 
for the laser focal spot radius of $r_\perp=2\, \mu$m,
the foil cannot be thinner than 
\begin{equation}
l_{0 \min}=
{N_e}/({\pi r_\perp^2 10^{24}\mbox{ cm}^{-3}}) \approx 1.6\, \mu{\rm m}. 
\end{equation}
The velocity of the preplasma expansion is equal to the 
ion acoustic velocity, $v_s=\sqrt{Z_i T_e/m_i}$
In the case of fully ionized aluminium, $Z_i=13$, $m_i\approx 27 m_p$,
we obtain $v_s \approx 1.2\times 10^7\,$cm/s for the electron temperature of $T_e=300\,$eV.
During one nanosecond the preplasma expands over the distance of $r_C=120\,\mu$m .
The critical surface is located at the distance of
$s_c \approx 20 \,\mu$m from the remaining portion of the foil,
assuming an exponential profile which is a typical case.


\subsection{Ion acceleration in preplasma}

When a laser pulse of femtosecond duration and of a petawatt level of power
(the {\it main pulse})
propagates through the {\it preplasma},
it undergoes two principal processes which
determine its evolution and, finally, 
the parameters of electromagnetic
radiation reaching the solid density foil and accelerating the ions.
The first process which we take into account is the {\it main pulse}
energy depletion due to the laser energy transformation to the energy
of fast electrons and ions. 
The second process is the self-focusing instability.

It is easy to obtain that the energy balance condition yields for the
depletion length, $l_{dep}$, of a relatively narrow laser pulse:
\begin{equation}
l_{dep}= (n_{cr}/n_e) a_0  l_{las}. 
\label{eq:depl}
\end{equation}
Here $l_{las}=c\tau_{las}$ and $a_0$ are the laser pulse length 
and dimensionless amplitude, respectively.

The laser pulse amplitude $a_0$ for given laser power ${\cal P}$ 
found from the self-focusing channel width is equal to \cite{SF-a-P} 
\begin{equation}
a_0= 
({\cal P}/{\cal P}_{cr})^{1/3} .
\label{eq:a1/3}
\end{equation}
Here ${\cal P}_{cr}=2 {\cal P}_{rel}(n_{cr}/n_e)\approx 17 (n_{cr}/n_e)$ GW
is the critical power of the relativistic self-focusing \cite{GZSUN}. A
laser pulse of the power $0.8\,$PW has the amplitude equal to $a_0=50$ which
corresponds to the radiation intensity of the order of $3\times
10^{21}$W/cm$^2$. Combining relationships (\ref{eq:depl}) and
(\ref{eq:a1/3}) we obtain that the length of the laser energy depletion in
the near-critical density plasma is
\begin{equation}
l_{dep}
\approx 45 l_{las}
\left(\frac{{\cal P}}{1\,\mbox{PW}} \right)^{1/3}
\left(\frac{10^{21}\,\mbox{cm}^{-3}}{n_e (\lambda [\mu\mbox{m}])^2} \right)^{4/3}
. 
\label{eq:ldep-a}
\end{equation}
If the laser pulse has a length $10\mu$m and power $0.8$ PW,
the energy depletion length for $n_e\approx 10 n_{cr}$ 
according to Eq. (\ref{eq:ldep-a}) is approximately equal to $19\mu$m. 
As we see, for the preplasma not thicker than $\approx
20\mu$m, the femtosecond PW power {\it main pulse} can deliver a substantial part
of its energy to the remaining portion of the foil,
while the pulse amplitude is $a_0=50$.
In this case, for a sufficiently thin remaining portion of the foil,
the ions can be accelerated by the laser radiation pressure in the Radiation Pressure
Dominated Acceleration (RPDA) regime \cite{RPDA, KAR}.

The key parameter characterizing the RPDA mechanism is the laser pulse fluence,
\begin{equation}
w(\psi)=\int_{-\infty}^{\psi}
(E^2(\psi')/2 n_0 l_0 m_i \omega^2 \lambda) d\psi'
\, ,
\label{eq:flu}
\end{equation}
written in the normalized form 
with $\psi = \omega t - \omega x(t)/c$, 
where $x(t)$ is the position of the foil remaining portion.
Below we shall use the total fluence $w=w(+\infty)$.

According to  Refs. \cite{RPDA, COMRE} (see also Ref. \cite{KLIMO, APLR}) 
in the nonrelativistic limit ($0<w\ll 1$), 
the maximum energy of ions accelerated in the RPDA regime scales as 
\begin{equation}
{\cal E}_i = 2 m_i c^2 w^2. 
\label{eq:Ei}
\end{equation}
According to the definition of the total fluence,
Eq. (\ref{eq:flu}) for $\psi=+\infty$,
the total number of accelerated ions, $N_i$,
is related to the laser pulse energy as 
\begin{equation}
N_i = {\cal E}_{las} / w m_i c^2 \, . 
\label{eq:Ei}
\end{equation}
Then the acceleration efficiency $\kappa_{eff}$
is proportional to the normalized total fluence 
\begin{equation}
\kappa_{eff} = N_i {\cal E}_i / {\cal E}_{las} = 2 w. 
\label{eq:keff}
\end{equation}

Using these relationships we obtain that in order to generate 
$N_i = 2\times 10^{11}$ protons per second with the energy of $250\,$MeV, 
one requires a 1 Hz laser with the pulse energy of ${\cal E}_{las}=30\,$J. 
For a 30 fs laser pulse duration this corresponds to the laser power of about 1 PW. 
The acceleration efficiency in this case is $\kappa_{eff}=0.27$.

The required thickness of the foil remaining portion, $\ell$,
or, more precisely, its surface density is determined by
the condition of target opaqueness \cite{RPDA, MACCHI, UNLIM, OPTIM},
necessary for RPDA:
\begin{equation}
a_0 < \epsilon_p.
\label{eq:epsp}
\end{equation}
The dimensionless parameter $\epsilon_p=2 \pi n e^2 \ell/m_e c \omega$ 
has been introduced in Ref. \cite{Vshivkov}.  
If a PW laser is focused to a spot with a size of $3\,\mu$m, 
i.e. its intensity is about $7\times 10^{21}$W/cm$^2$,
for the fully ionized aluminium solid density, $n_e = 7.8\times 10^{23}$cm$^{-3}$,
the foil remaining portion thickness should be not thinner than the order of $l=32\,$nm.

Since one of the most important features of the RPDA acceleration regime is
the laminarity of the accelerated ion beams, 
by using laser pulses with a super-Gaussian transverse profile
and 
narrow aperture collimators for ions,
one can obtain a narrow energy spectrum ion beam.


\section{Hydrodynamic Modelling of the Preplasma Formation}
\label{sec:HD}

We have run several hydrodynamic simulations all giving similar features \cite{UTSUMI, SUNAHARA,OHNISHI}.
In order to simulate the interaction of the {\it prepulse} with the solid
target we have chosen results from the two-dimensional Arbitrary Lagrangian Eulerian (ALE)
radiation hydrodynamics code STAR2D as a representative result \cite{SUNAHARA}.

\subsection{Governing Equations}

The governing equations used in the code are \cite{SUNAHARA}:
\begin{align}
\frac{d\rho}{dt} = & -\rho\vec\nabla\cdot\vec v, \label{mass conservation} \\
\rho\frac{d\vec v}{dt} = & -\vec\nabla (p+q), \label{force} \\
\rho c_{\nu i}\frac{dT_i}{dt} = & - (p_{THi} + q)\vec\nabla\cdot\vec v + \vec\nabla\cdot(\kappa_i\vec\nabla T_i)\nonumber \\
                                & + \alpha(T_e-T_i), \label{ion temp} \\
\rho c_{\nu e}\frac{dT_e}{dt} = & - p_{THe}\vec\nabla\cdot\vec v + \vec\nabla\cdot(\kappa_e\vec\nabla T_e) \nonumber \\
                                & - \alpha(T_e-T_i) + Q_L + Q_r, \label{electron temp} \\
\rho \frac{d}{dt}\left(\frac{E^\nu}{\rho}\right) = {}& \Omega\cdot(D^\nu\nabla E^\nu) + 4 \pi\eta^\nu - c\chi^\nu E^\nu,
\label{rt}
\end{align}
where $\vec v$ is the velocity, $\rho$ is the mass density, 
$p$ is the total pressure $p=p_e+p_i$ which is the sum of the 
electron pressure $p_e$ and ion pressure $p_i$,
$q$ is the artificial viscosity used for regularization of the shock wave fronts \cite{RICHTMYER},  
$p_{THi}\equiv T_i (\partial p_i/\partial T_i)$, 
$p_{THe}\equiv T_i (\partial p_e/\partial T_e)$,
$T_i$ and $T_e$ are the ion and electron temperature, respectively,
$c_{\nu i}$ and $ c_{\nu e}$ are the specific ion heat and electron heat, respectively,
$\kappa_i$ and $\kappa_e$ are the ion and electron conductivities, respectively, 
including the flux-limited Spitzer-Harm model.
The term $\alpha(T_e-T_i)$ incorporates electron-ion temperature relaxation 
where $\alpha$ is determined from the Spitzer relaxation time,
$Q_L$ is the source term from laser heating of electrons where the 
laser absorption process is assumed to be inverse-bremsstrahlung and 
the laser propagation is calculated by ray-tracing,
$Q_r$ is the heating term due to radiation.
In the radiation transport equation, Eq. (\ref{rt}),
$E^\nu$ is the photon energy density at energy $h\nu$,
$\Omega$ is the x-ray propagation direction, 
$\eta^\nu$ is the emissivity,
$\chi^\nu$ is the attenuation coefficient, and
$D^\nu$ is a diffusion coefficient defined by 
$D^\nu \equiv c/(3\chi^\nu + c |\nabla E^\nu|/E^\nu)$.  
More details can be found in \cite{SUNAHARA}.
The  equations of state (EOS) used in 
Eqs. (\ref{force}),(\ref{ion temp}), and (\ref{electron temp}) are
that of Ref. \cite{RAY}.   
Details and advantages of using this EOS versus other EOS' will be presented elsewhere. 
 
\subsection{Parameters and Results}

We model the {\it prepulse} as a square pulse of duration 3 ns 
and transverse Gaussian profile with a wavelength of 1.06$\,\mu$m.
The peak intensity on target is $2 \times 10^{11}$ W/cm$^2$.
The full width at half maximum (FWHM) focus spot size is taken to be 2 $\mu$m.
We get the {\it prepulse} energy of $2.7 \times 10^{-5}$ J.
Taking the {\it main pulse} energy to be 20 J
the energy contrast, $C_{\cal E}$ defined by Eq. \ref{eq:CE}, is
 $7.4\times10^{5}$.

\begin{figure}[tbph]
\includegraphics[width=0.99\columnwidth]{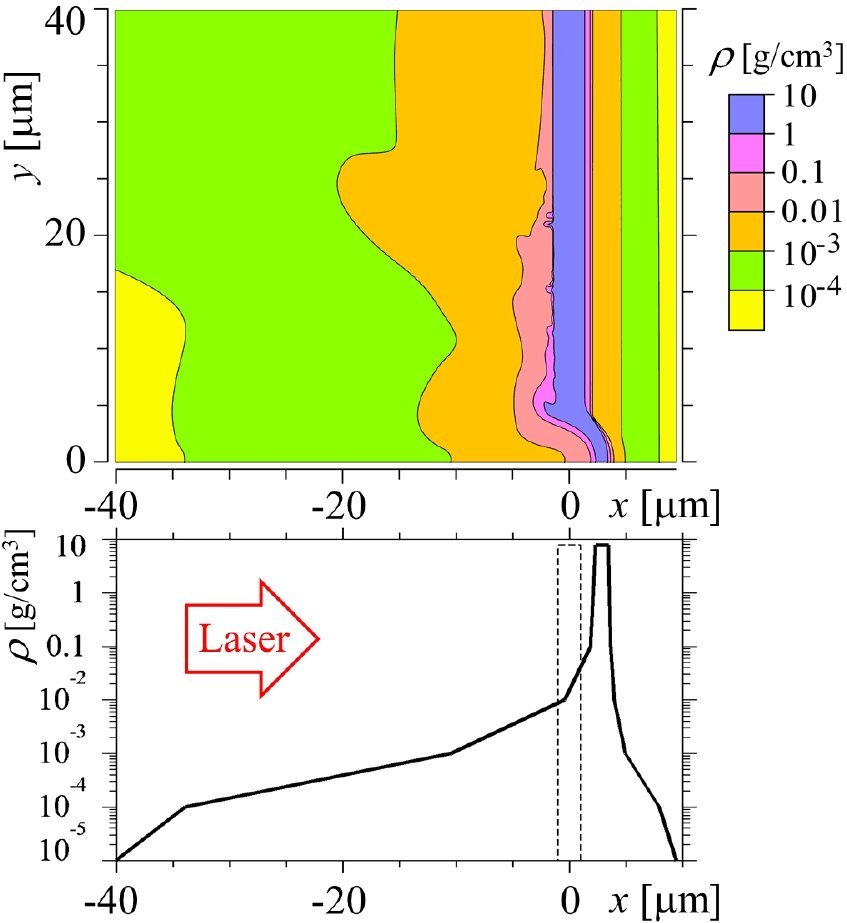}
\caption{\label{FIG3}
Density of iron target (g/cm$^3$) after 3 ns of irradiation
by the {\it prepulse} with a 2 $\mu$m spot FWHM size,
and intensity $2 \times 10^{11}$ W/cm$^2$.
Below is the density along $x$-axis (solid curve) and
the initial target profile (dashed curve).
}
\end{figure}

Assuming that the {\it main pulse} is Gaussian and 
has the FWHM duration of 30 fs with respect to intensity,
the intensity contrast, $C_I$ defined by Eq. \ref{eq:CI},
is $6.9 \times 10^{10}$.
The target material is iron with a thickness of 2 $\mu$m
and the initial density of 7.8 g/cm$^3$.
The simulation box size is 80 $\mu$m in the laser propagation direction 
and 60 $\mu$m in the transverse direction.
In Fig. \ref{FIG3}
the laser is propagating from the left to the right perpendicular to the target.  

Figure \ref{FIG3} shows the density of the iron target
near the laser spot 
after being irradiated for 3 ns 
by the {\it prepulse}.
It can be seen that the target has become curved away from the direction of the laser propagation.
The target has moved approximately 3.3 $\mu$m from its initial position.
The portion of the target where the {\it prepulse} intensity is greatest 
has thinned relative to its initial 2 $\mu$m thickness.
The bottom of Fig. \ref{FIG3} shows a line profile at $y=0$ of the density of 
the aluminum target, at the same time as the top of Fig. \ref{FIG3}. 
The shift in the direction of the laser propagation can be seen.


\section{Multiparametric Particle-In-Cell Simulations of the Ion Acceleration}
\label{sec:PIC}


Using the results of dissipative hydrodynamic simulations 
here we investigate the interaction
of the high intensity femtosecond portion of the laser pulse (the {\it main pulse})
with the Al (aluminum) foil
modified by the low intensity nanosecond portion (the {\it prepulse}).
We search for 
the dependence of the maximum ion energy 
on the laser beam and target properties.

\subsection{Plasma profile}
As we saw in the previous section,
when a metallic foil is irradiated by a {\it prepulse},
a portion of the foil evaporates forming a {\it preplasma}, Fig \ref{fig:prep-plat-skirt},
while the remaining portion with a flat-top profile ({\it plateau})
is bent due to a recoil with the transverse scale 
of the order of the laser pulse spot size, Fig. \ref{FIG3}.
On the back side of the foil a relatively thin and low density plasma layer
({\it skirt}) is formed.

A sufficiently thin foil can be completely disintegrated.
While there are several mechanisms of ion acceleration by intense femtosecond pulses
in underdense or near-critical plasma \cite{MVA,MATSU,YOGO,Fukuda},
here we consider the case where a significant portion of the foil
has a solid density at the arrival of the {\it main pulse}
which may provide conditions for the RPDA and directed-Coulomb explosion regimes.

Our hydrodynamics simulations for micron-thick foils
show that the profiles of the {\it preplasma} and {\it skirt} 
are nearly exponential up to the critical density,
while the critical surface (for the resulting maximum ionization state)
is not farther than a few microns away from the {\it plateau}.
Assuming a typical shape of the {\it preplasma} and varying its scale
one can cover a wide range of possible {\it preplasma} profiles
without a loss of qualitative fitness of the results.
We model the plasma (electron) density by the formulae
\begin{align}
\rho(x,y) =
{}& n_1 \Lambda\left( \textstyle\frac{x +l_0/2 - s_f B(y)}{L_{p}}, \frac{y(1-B(y))}{L_{p}} \right)
\label{eq:foil}\\
{}& +n_0 \Pi\left( \textstyle\frac{x +l_0/2 - s_f B(y)}{L_{p}}, \frac{x-l_0/2 - s_b B(y)}{L_{p}} \right)
\nonumber \\
{}& +n_c \Lambda\left( \textstyle\frac{-x+l_0/2 + s_b B(y)}{L_{pb}}, \frac{y(1-B(y))}{s_\perp} \right)
\nonumber , \\
\Lambda(\xi,\eta) = {}& 10^{-\sqrt{\xi^2+\eta^2}} \theta(-\xi) , \label{eq:slope} \\
\Pi(\xi_1,\xi_2)  = {}& \theta(\xi_1) \theta(-\xi_2) , \label{eq:top} \\
             B(y) = {}& e^{-y^2/s_\perp^2}
,
\end{align}
where $\theta$ is the Heaviside step function,
$\theta(\xi)=0$ for $\xi<0$ and $\theta(\xi)=1$ for $\xi>0$.
The resulting density profile is shown in Fig. \ref{fig:model}(a).
The unperturbed part of the foil
with the (electron) density $n_0$
occupies the interval of $|x| < l_0/2$ for $|y|>s_\perp$,
where $l_0$ is the initial foil thickness.
The remaining portion of the foil, {\it plateau}, 
occupies the interval of $|x - s_b B(y)| < l_0/2$ for $|y|<s_\perp$,
where the function $B(y)$ and parameters $s_\perp$, $s_f$ and $s_b$ describe bending.
The {\it plateau} (minimum) thickness is $\ell = l_0+s_b-s_f$, $0<\ell<l_0$.
The {\it preplasma} with the maximum (electron) density of $n_1$
is characterized by the scale-length $L_p$,
the distance at which the plasma density drops 10 times.
The {\it skirt} at the back side of the foil 
has a longitudinal and transverse scale-length of 
$L_{pb}$ and $s_\perp$, respectively.
Its maximum (electron) density is assumed to be equal to the critical density, $n_c$,
for the maximum ionization state resulting from the hydrodynamic simulations.

We also model water contamination usually present at the back side of the foil.
This additional layer with the thickness of $l_{0H}$ (typically 10 nm)
is adjacent to the back side of the foil. 
Its density is a combination of 
the flat-top profile Eq. (\ref{eq:top})
and the exponential profile Eq. (\ref{eq:slope}),
in the same fashion as the {\it skirt} density [the last term in Eq.(\ref{eq:foil}),
where the first argument is replaced by $(-x+l_0/2 + (s_b + 2.5 l_{0H}) B(y))/L_{pb}$].
The longitudinal and transverse scale-length of the exponential profile
are the same as for the {\it skirt}, i.e. $L_{pb}$ and $s_\perp$, respectively.
The maximum (electron) density in this layer is $n_H$.

The number of particles ablated from the foil
is equal to the {\it preplasma} density integral, neglecting the {\it skirt} integral.
Assuming rotational symmetry, we have
\begin{align}
{}&
2\pi n_0 \int \Pi y dx dy = 2\pi n_1 \int \Lambda y dx dy , \\
{}&
\pi n_0 (s_f-s_b) s_\perp^2 = \pi n_1 L_p^3 {\cal I}(s_\perp/L_p), \label{eq:N-ablated} \\
{}&
{\cal I}(\mu) = \int_0^{+\infty}d\eta \int_{-\infty}^{+\infty}
10^{-\sqrt{\xi^2+\eta(1-\exp(-\eta/\mu^2))^2}}d\xi
,
\end{align}
where the function ${\cal I}(\mu)$ can be approximated on the interval $0.01<\mu<1$ by
$
{\cal I}(\mu) = 0.328 + \mu^4 (6-7.7\mu+4.48\mu^2-1.05\mu^3)/(1+7.75\mu)
$
with the relative error of 0.1\%.
Since $s_f-s_b = l_0-\ell$,
this relation determines $n_1$ for given $\ell$, $L_p$ and $s_\perp$:
\begin{equation}
n_1 = n_0 \frac{l_0-\ell}{L_p} \frac{s_\perp^2/L_p^2}{{\cal I}(s_\perp/L_p)}
.
\end{equation}
We assume that $n_1\le n_0$.
The amount of ablated material is proportional to $l_0-\ell$
and does not depend on $L_p$, Eq. (\ref{eq:N-ablated}).

The critical surface at which the {\it preplasma} (electron) density equals critical density, $n_c$,
is determined by the equation $\rho(x,y) = n_c$.
On the rotational symmetry axis, this reads $n_1 10^{-s_c/L_p} = n_c$,
where $s_c$ is the distance between the critical surface
and the {\it plateau}, 
\begin{equation} \label{eq:critsurf}
s_c = L_p \lg \left( \frac{n_0}{n_c} 
\frac{l_0-\ell}{L_p} \frac{s_\perp^2/L_p^2}{{\cal I}(s_\perp/L_p)} \right)
.
\end{equation}
We note that the expressions for $n_1$ and $s_c$ 
do not explicitly depend on the longitudinal shift of the bent foil,
described by parameters $s_f$ and $s_b$.

The electron density obtained from the hydrodynamics simulations
typically corresponds to a relatively low degree of ionization of atoms
constituting the foil.
For example, in the previous section the {\it prepulse}
with the intensity of $10^{11}$ W/cm$^2$
produces preplasma with Al$^{+3}$. 
For the density of aluminum under the normal conditions,
3 electrons per atom give the electron density about $104 n_{cr}$
while a full ionization implies $450 n_{cr}$,
where $n_{cr} = \pi/(r_e \lambda^2) = 1.74\times10^{21}$ cm$^{-3}$,
$r_e$ is the classical electron radius.
A 10 J, 30 fs {\it main pulse} focused to a 3 $\mu$m spot
is strong enough to fully ionize Al atoms,
also aided by a collective electrostatic field in the preplasma.
Therefore, in the simulations we assume that the plasma is fully ionized.
However, for experimental diagnostics it is useful to know
the location of the critical surface just after the nanosecond pulse,
$s'_c$.
It is given by the same formula as Eq. (\ref{eq:critsurf}),
with $n_0$ replaced by the (electron) density corresponding to
a lower degree of ionization.
For both assumptions the critical surface location
is shown in Fig. \ref{fig:model}(b).

\subsection{Simulation setup}
We use the particle-in-cell (PIC) code REMP
in a two-dimensional (2D) configuration
for the multi-parametric (MP) study, where
several tasks with different sets of initial parameters
are simultaneously performed on a supercomputer.
The simulation box size is $130\lambda \times 72\lambda$.
Boundary conditions are absorbing for electromagnetic field and quasiparticles.
The mesh size is $\lambda/16$.
The time step is $0.011\lambda/c$.
The total number of quasi-particles representing electrons, protons
and O$^{+8}$ and Al$^{+13}$ ions
is $4.4\times 10^8$.

\begin{figure}[tbph]
\includegraphics[width=0.75\columnwidth]{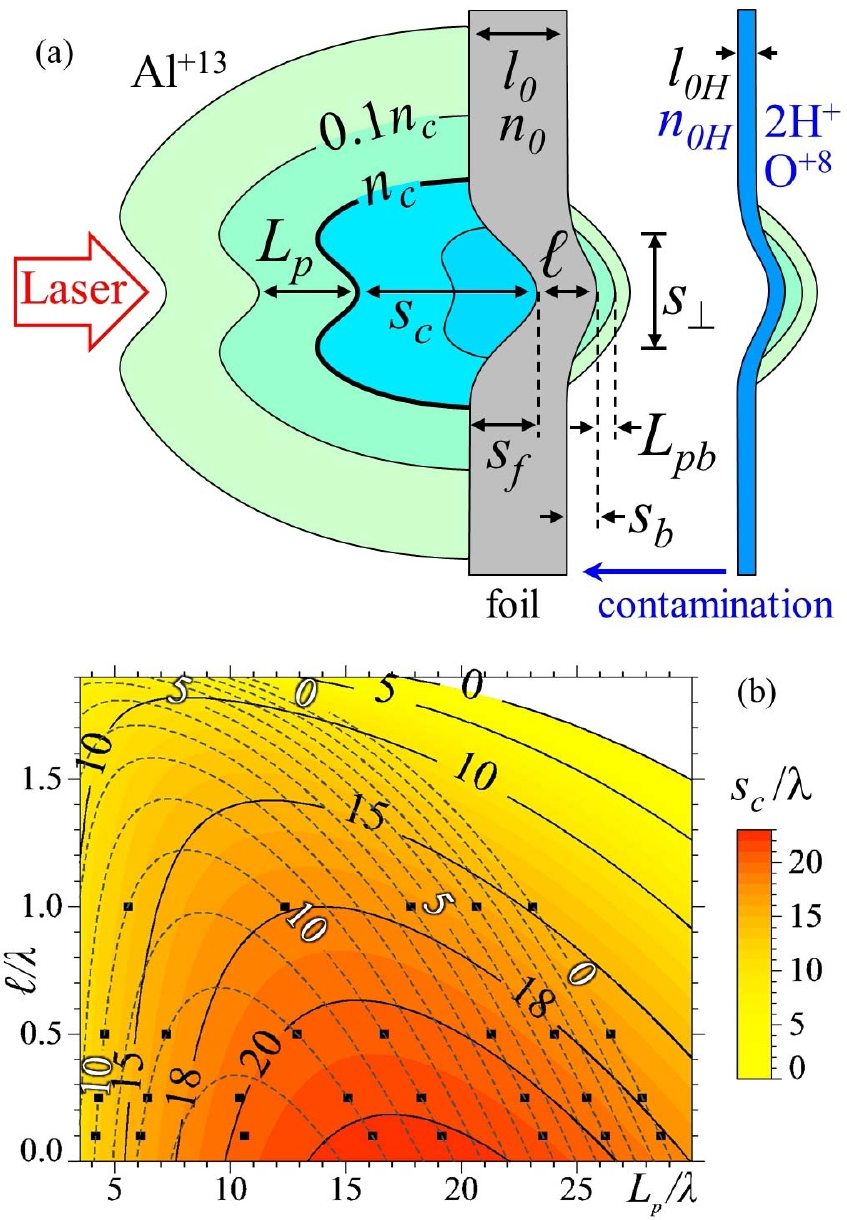}
\caption{\label{fig:model}
(Color online).
(a) Preplasma model used for multi-parametric PIC simulations. 
The parameters $L_p$ and $\ell$ vary while other parameters are fixed.
The water contamination layer is attached to the back side of the foil.
(b) The choice of parameters $L_p$, $\ell$ shown by squares.
Solid lines: the distance from the {\it plateau}
to the critical surface, $s_c$, in units of $\lambda$ for fully ionized Al$^{+13}$;
dashed lines: a similarly defined distance, $s'_c$, for Al$^{+3}$.
}
\end{figure}

We fix the following parameters of the target,
typically observed in hydrodynamic simulations 
in a wide range of the {\it prepulse} parameters.
The unpertubed portion of the aluminum foil
has the thickness of $l_0 = 2\lambda$.
Its center is placed at the distance of
$50\lambda$ from the left boundary of the simulation box.
Transverse scale of bending is $s_\perp = 3\lambda$
(this is approximately equal to the focal spot size of a laser beam
focused by a f/3 system).
The foil back side shift (due to bending) is $s_b = 1.25\lambda$.
The {\it skirt} has a longitudinal scale-length of $L_{pb}=0.5\lambda$,
and transverse scale-length of $s_\perp$.
The water layer thickness is $l_{0H} = 0.01\lambda$.
The {\it plateau} (electron) density is $n_0 = 450 n_{cr}$,
the water layer (electron) density is $n_H = 154 n_{cr}$.
The {\it preplasma} scale-length, $L_p$, varies from $4\lambda$ to $\sim 29\lambda$.
The {\it plateau} thickness, $\ell$, varies from $0.1\lambda$ to $1\lambda$.
The choice of parameters $L_p$, $\ell$ is shown in Fig. \ref{fig:model}(b).
We also consider an ideal case where
a ``clean'' femtosecond laser pulse interacts with an unperturbed foil,
i.~e. formally $L_p=0$.

The {\it main pulse} is approximated by a Gaussian beam
with the wavelength of $\lambda=0.8\ \mu$m,
linearly polarized along the $y$-axis.
Originating from $x=0$, it is focused with the f-number of f/3
normally onto the plane at the front surface of the unperturbed foil.
In vacuum, its focal spot full-width-at-half-maximum (FWHM) size 
would be $3\lambda$ (with respect to intensity).
The {\it main pulse} duration (FWHM with respect to intensity) is 30 fs $= 11.24\lambda/c$.
The {\it main pulse} energy takes the values of 4 J, 10 J, and 20 J.

\begin{figure}[tbph]
\includegraphics[width=0.8\columnwidth]{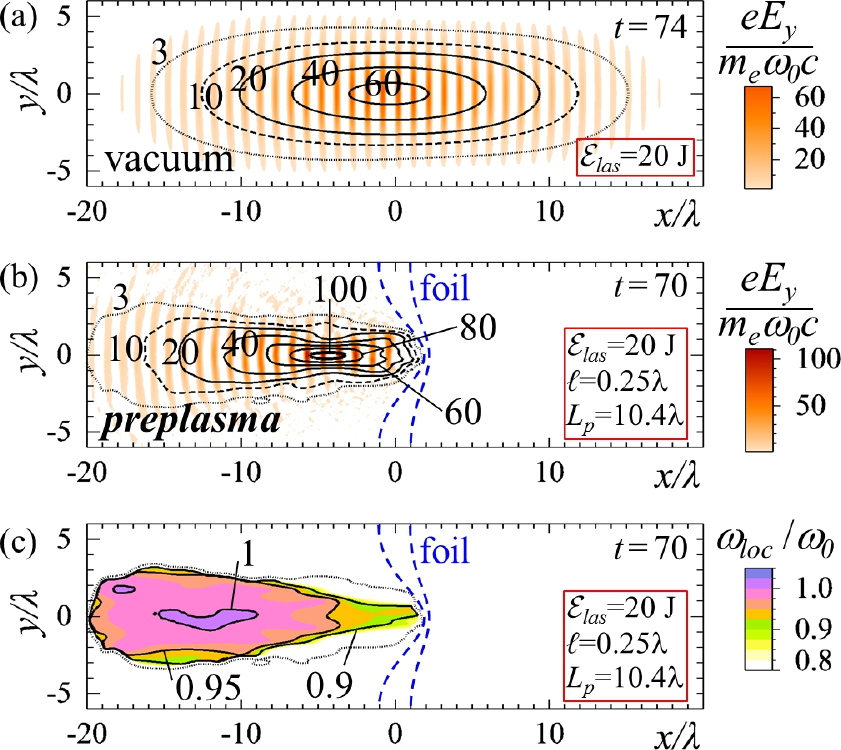}
\caption{\label{fig:focus}
(Color online).
The {\it main pulse} near its focal plane 
at the moment of achieving maximum intensity
(a) in vacuum, and (b) in {\it preplasma}.
(c) The local carrier frequency of the {\it main pulse}
for the latter case.
}
\end{figure}

\begin{figure}[tbph]
\includegraphics[width=0.75\columnwidth]{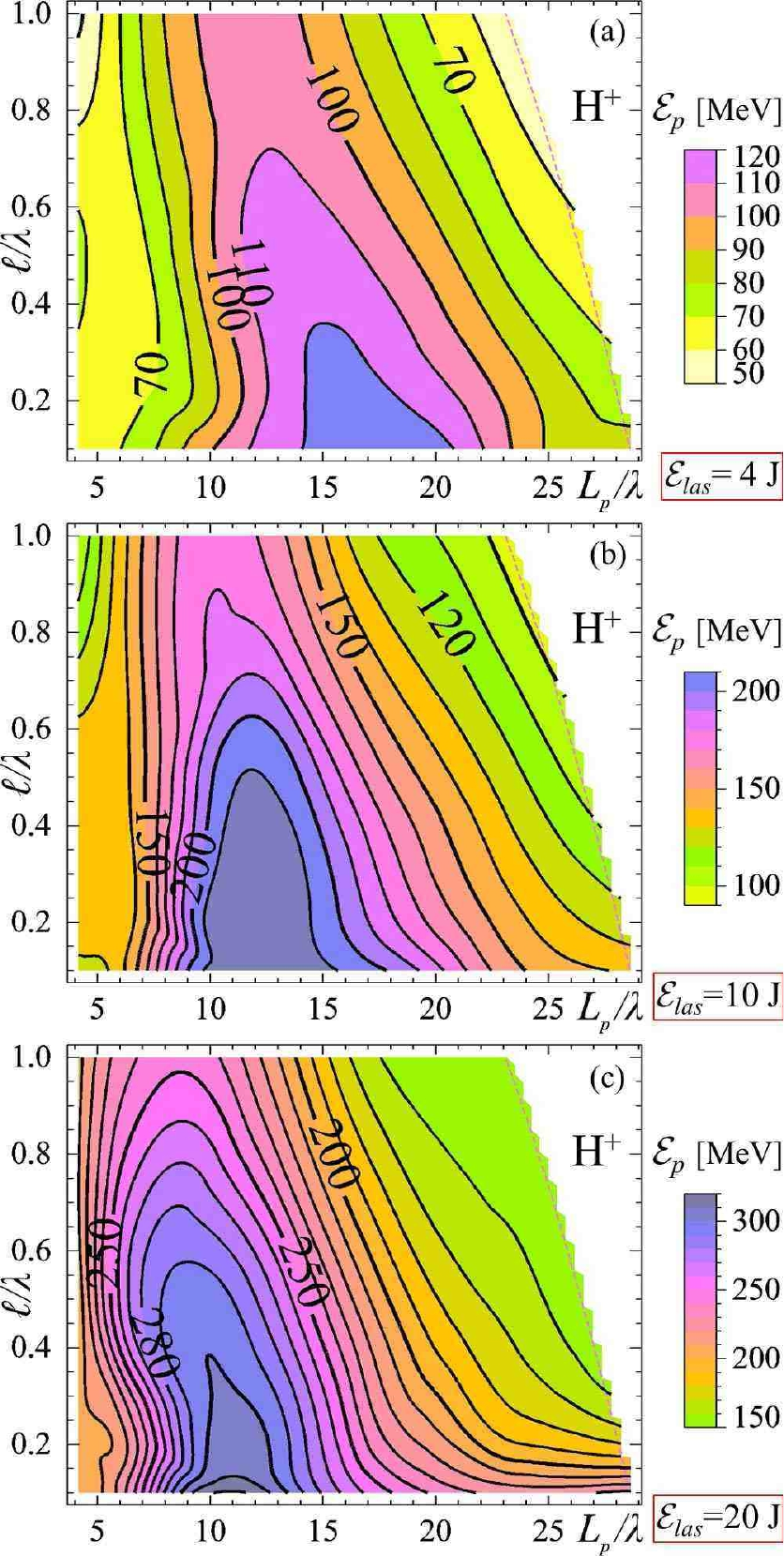}
\caption{\label{fig:en-H}
(Color online).
Maximum proton energy in MeV (colorscale, solid curves) 
as a function of $L_p$ and $\ell$
at $t=140 \lambda/c$
for the {\it main pulse} energy of (a) 4 J, (b) 10 J, and (c) 20 J.
}
\end{figure}

\begin{figure}[tbph]
\includegraphics[width=0.75\columnwidth]{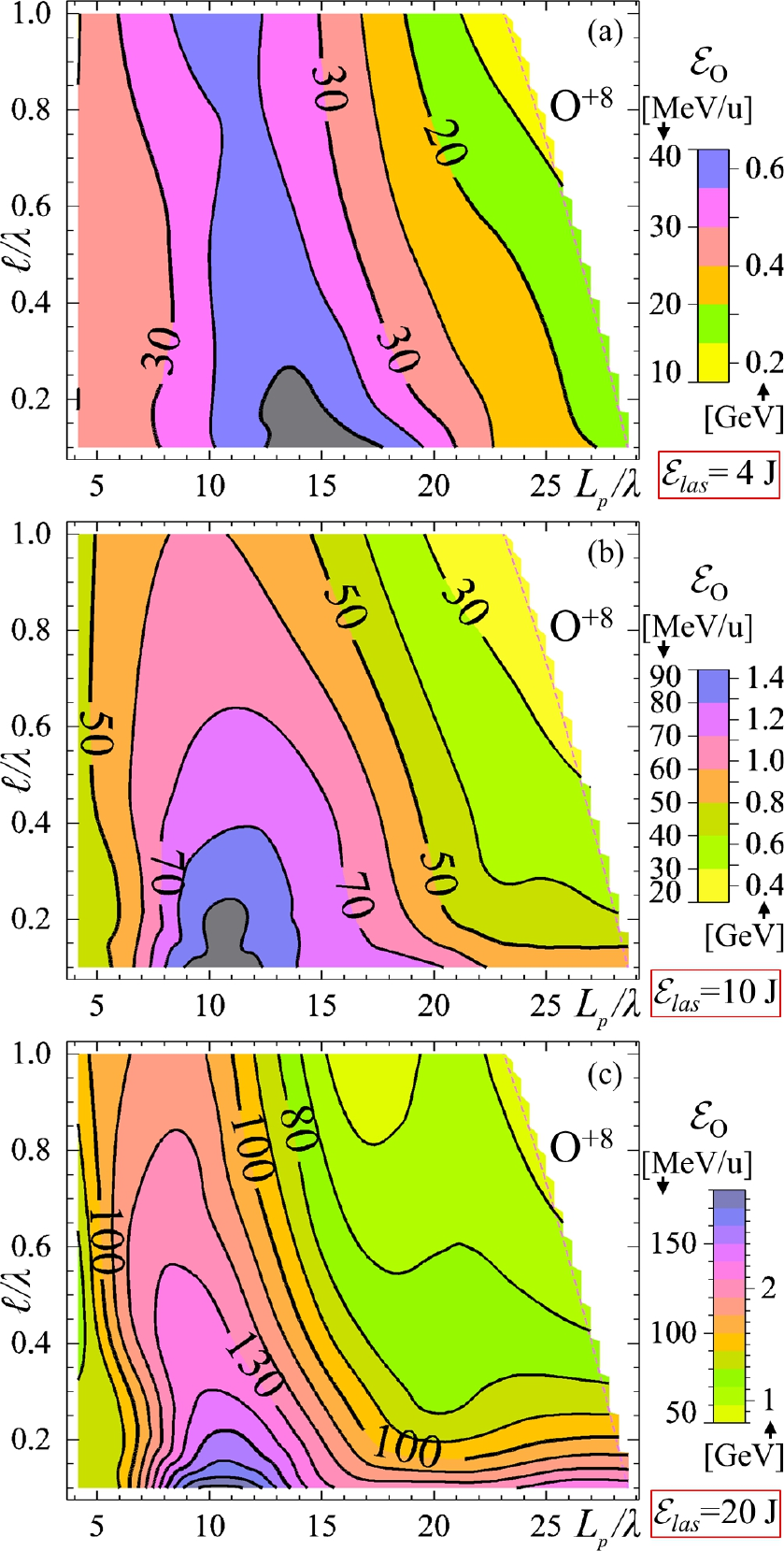}
\caption{\label{fig:en-O}
(Color online).
Maximum O$^{+8}$ energy in MeV/nucleon (colorscale, solid curves) 
as a function of $L_p$ and $\ell$
at $t=140 \lambda/c$
for the {\it main pulse} energy of (a) 4 J, (b) 10 J, and (c) 20 J.
The values on the right from the colorscale correspond to the total ion energy in GeV.
}
\end{figure}

\begin{figure}[tbph]
\includegraphics[width=0.75\columnwidth]{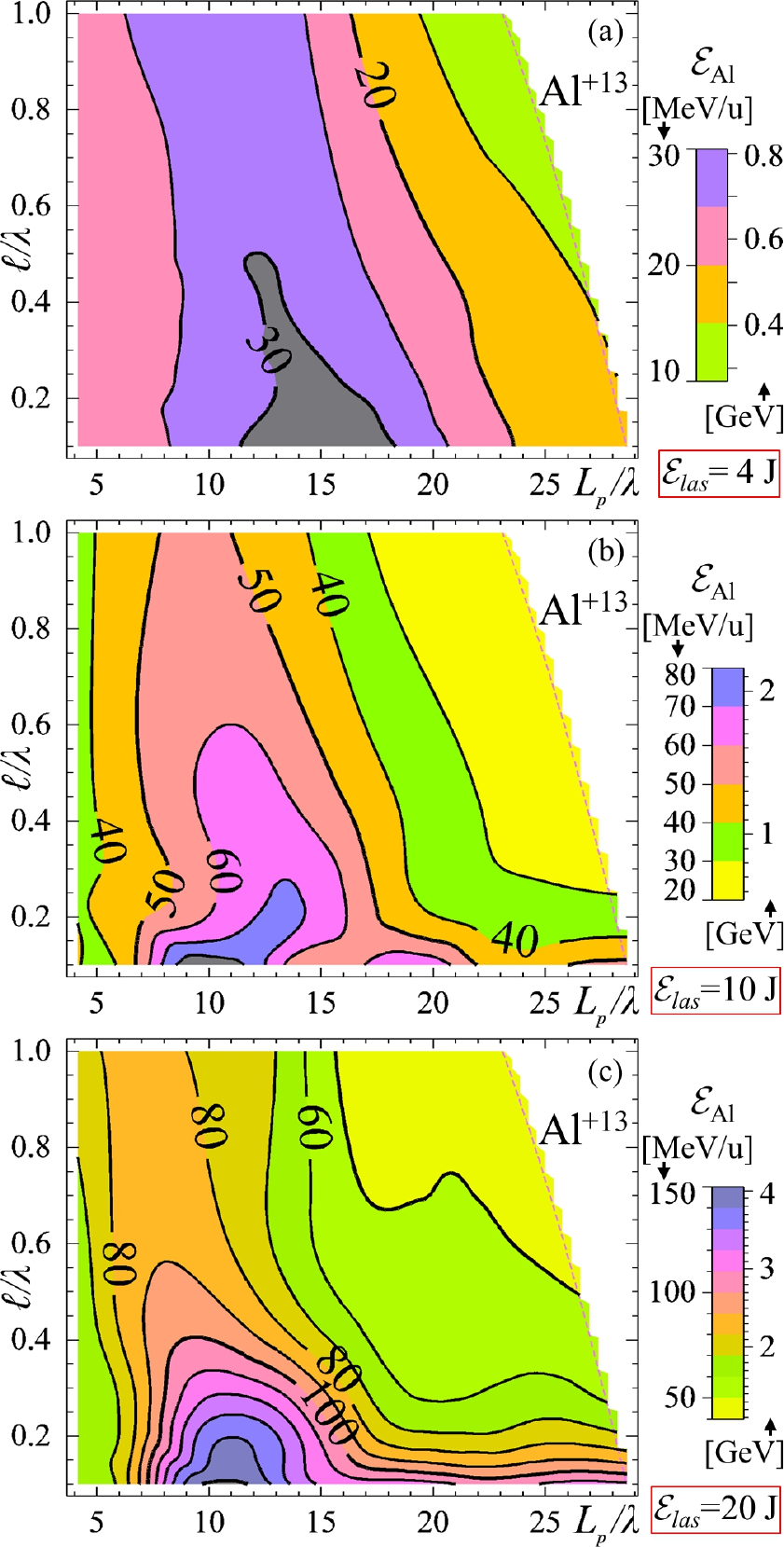}
\caption{\label{fig:en-Al}
(Color online).
Maximum Al$^{+13}$ energy  in MeV/nucleon (colorscale, solid curves) 
as a function of $L_p$ and $\ell$
at $t=140 \lambda/c$
for the {\it main pulse} energy of (a) 4 J, (b) 10 J, and (c) 20 J.
The values on the right from the colorscale correspond to the total ion energy in GeV.
}
\end{figure}

\begin{figure}[tbph]
\includegraphics[width=0.75\columnwidth]{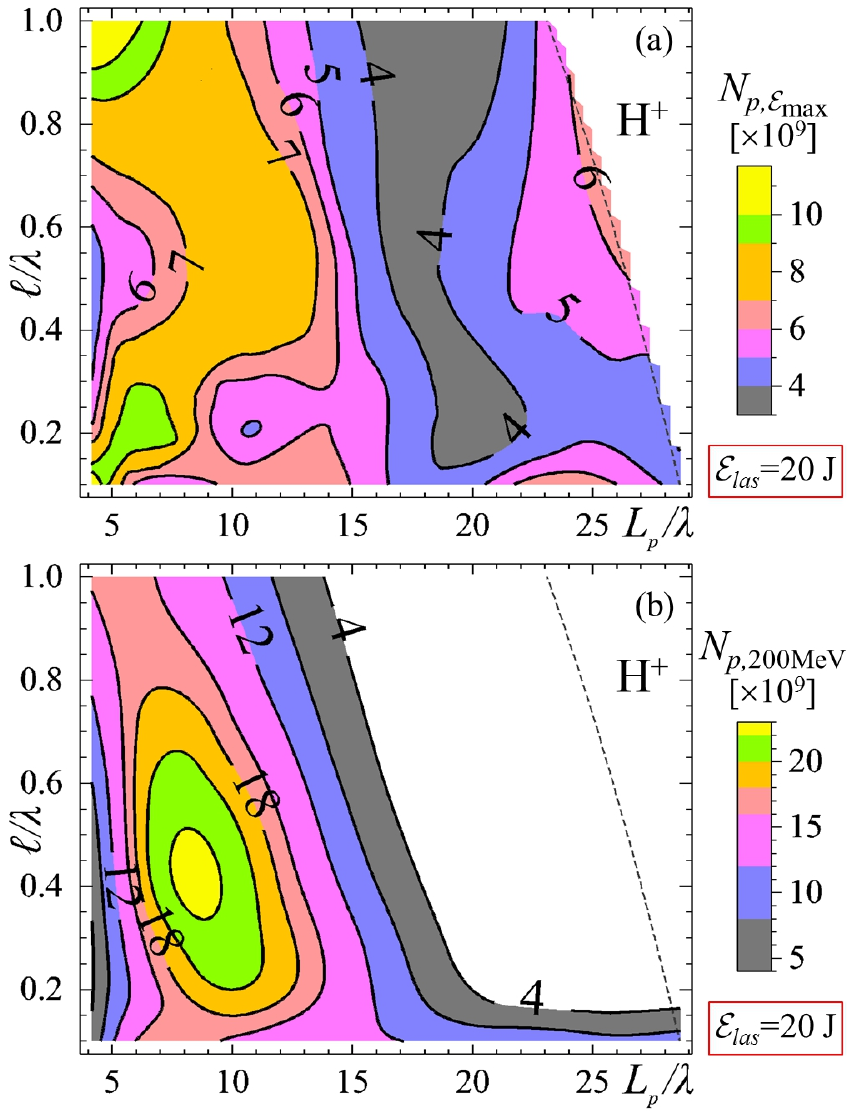}
\caption{\label{fig:num-H}
(Color online).
The proton number with the energy greater than 90\% of maximum (a),
and with the energy in the interval of $200\pm 10$ MeV (b)
for the {\it main pulse} energy of 20 J
at $t=140 \lambda/c$.
}
\end{figure}

\begin{figure}[tbph]
\includegraphics[width=0.8\columnwidth]{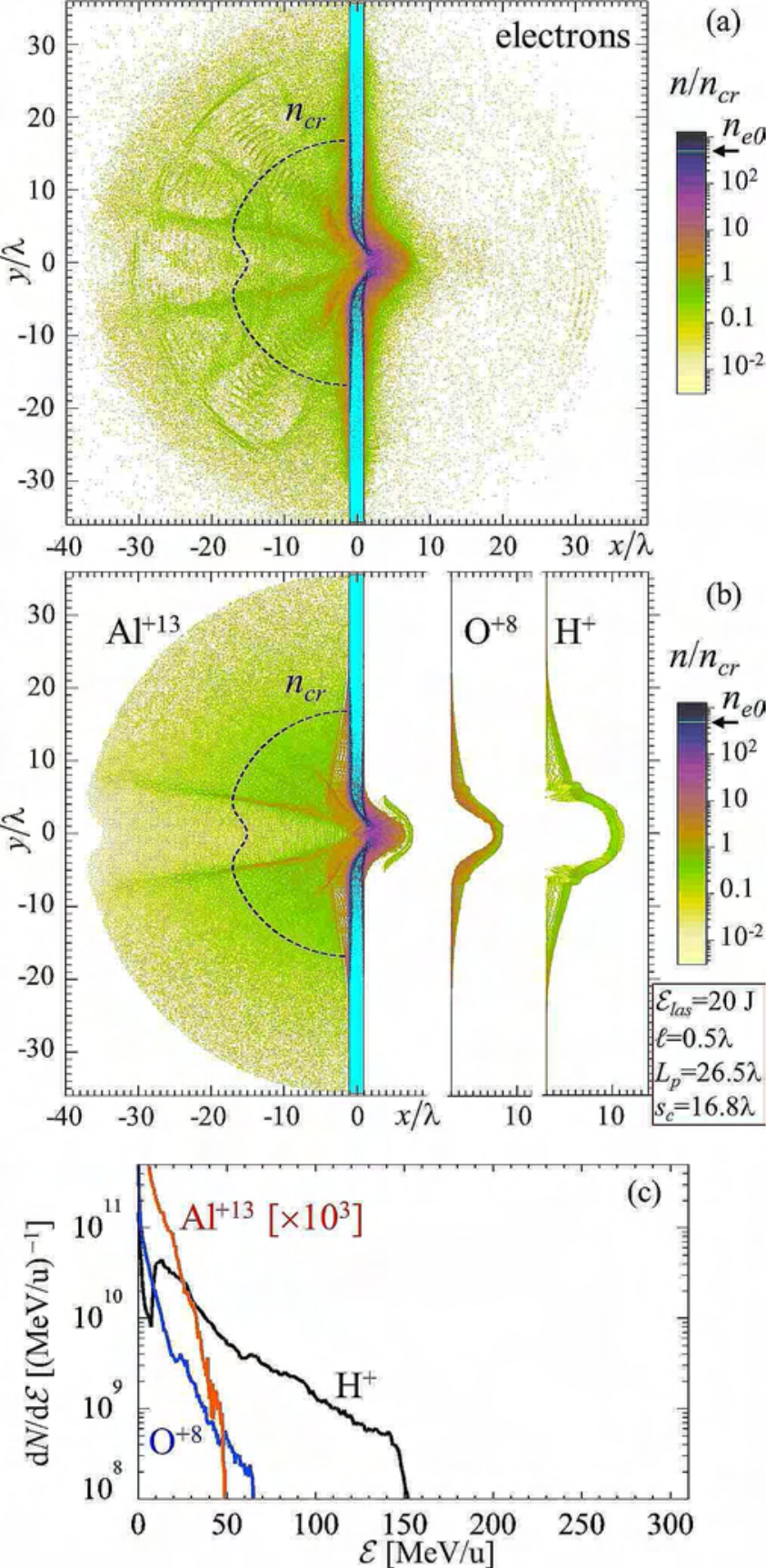}
\caption{\label{fig:150-dens}
(Color online).
Electron (a) and ions (b) densities  
for $\ell=0.5\lambda$, $L_p=26.5\lambda$ (for which $s_c=16.8\lambda$)
obtained with the 20 J {\it main pulse}
at $t=100 \lambda/c$.
(c) Energy spectra for ions at $t=140 \lambda/c$.
The distribution for Al$^{+13}$ is multiplied by $10^3$.
}
\end{figure}

\begin{figure}[tbph]
\includegraphics[width=0.8\columnwidth]{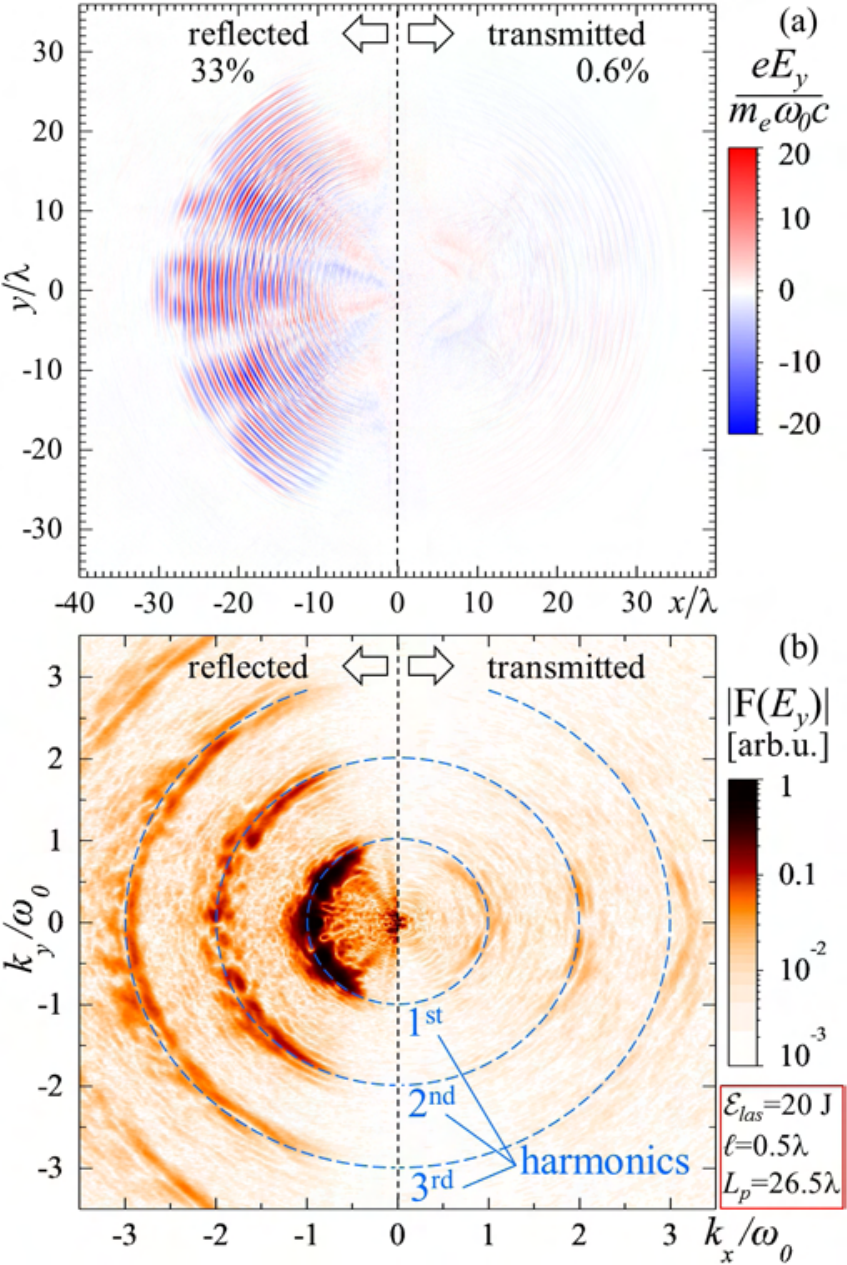}
\caption{\label{fig:150-ey}
(Color online).
Electric field $E_y$ component (a)
and the absolute value of its fast fourier transform (b)
revealing harmonics in the reflected and transmitted radiation;
observed in the same case as Fig. \ref{fig:150-dens}
at $t=100 \lambda/c$.
The reflected {\it main pulse} energy 
is 33\%, while 0.6\% is transmitted and the rest is absorbed.
}
\end{figure}

\begin{figure}[tbph]
\includegraphics[width=0.8\columnwidth]{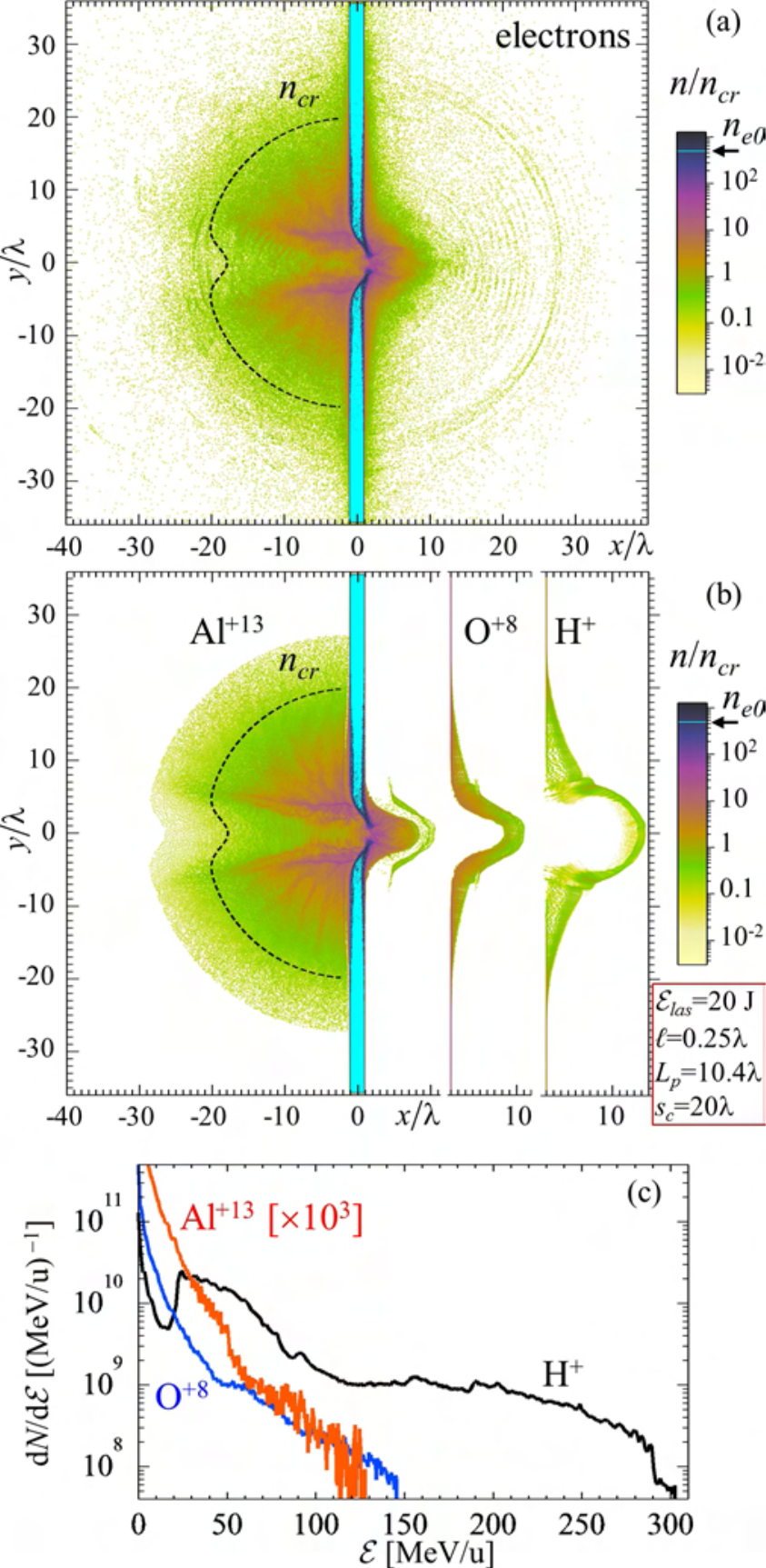}
\caption{\label{fig:300-dens}
(Color online).
Electron (a) and ions (b) densities  
for $\ell=0.25\lambda$, $L_p=10.4\lambda$ (for which $s_c=20\lambda$)
obtained with the 20 J {\it main pulse}
at $t=100 \lambda/c$.
(c) Energy spectra for ions at $t=140 \lambda/c$.
The distribution for Al$^{+13}$ is multiplied by $10^3$.
}
\end{figure}

\begin{figure}[tbph]
\includegraphics[width=0.8\columnwidth]{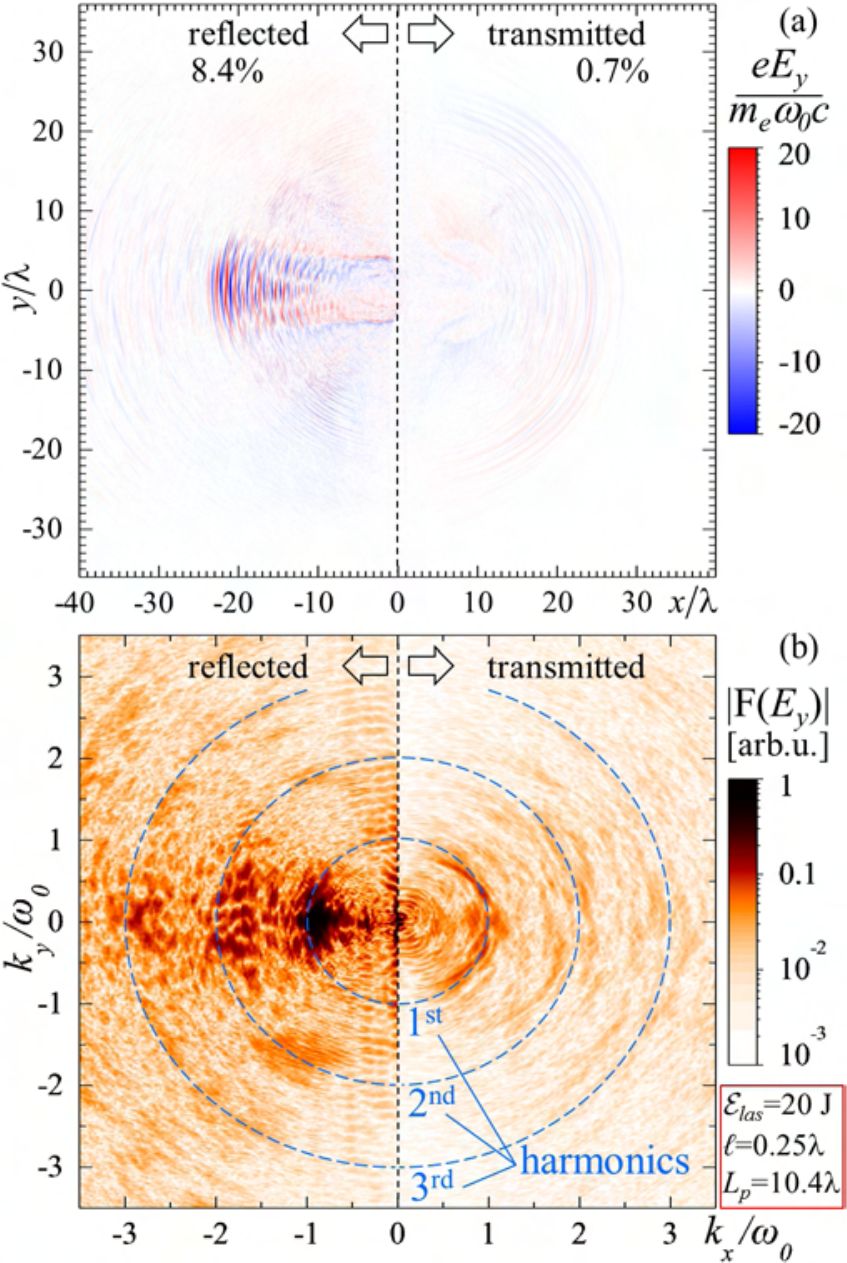}
\caption{\label{fig:300-ey}
(Color online).
Electric field $E_y$ component (a)
and the absolute value of its fast fourier transform (b)
revealing harmonics in the reflected and transmitted radiation;
observed in the same case as Fig. \ref{fig:300-dens}
at $t=100 \lambda/c$.
The reflected {\it main pulse} energy 
is 8.4\%, while 0.7\% is transmitted and the rest is absorbed.
}
\end{figure}

\begin{figure}[tbph]
\includegraphics[width=0.75\columnwidth]{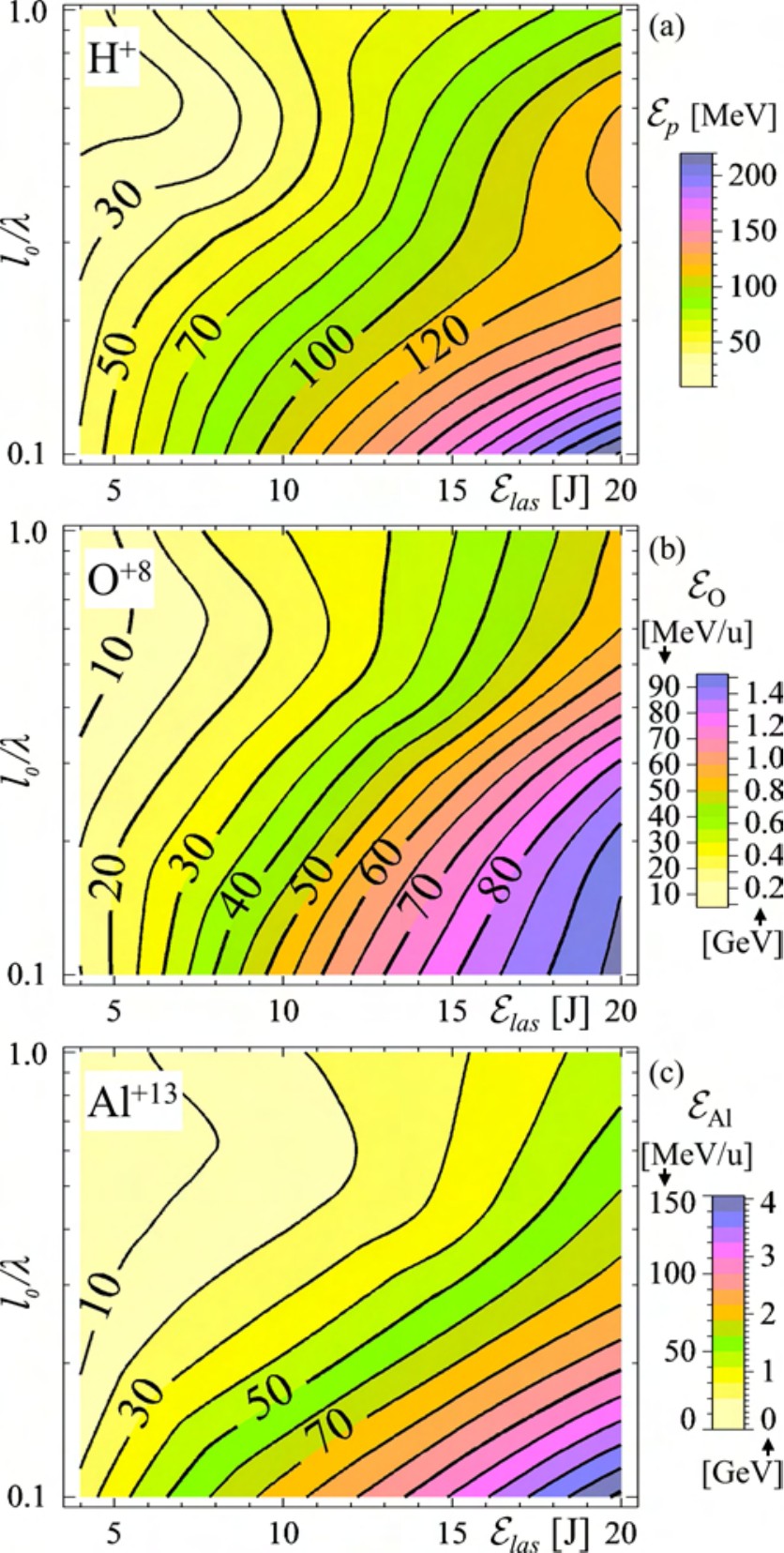}
\caption{\label{fig:en-H-clean}
Maximum energy of (a) protons in MeV,
and of (b) O$^{+8}$ and (c) Al$^{+13}$ ions in MeV/nucleon (colorscale, solid curves) 
as a function of the foil thickness, $\ell$,
and the {\it main pulse} energy, ${\cal E}_{las}$,
in the case whithout {\it preplasma} (formally $L_p=0$) 
at $t=140 \lambda/c$.
The values on the right from the colorscale correspond to the total ion energy in GeV.
}
\end{figure}

\begin{figure}[tbph]
\includegraphics[width=0.8\columnwidth]{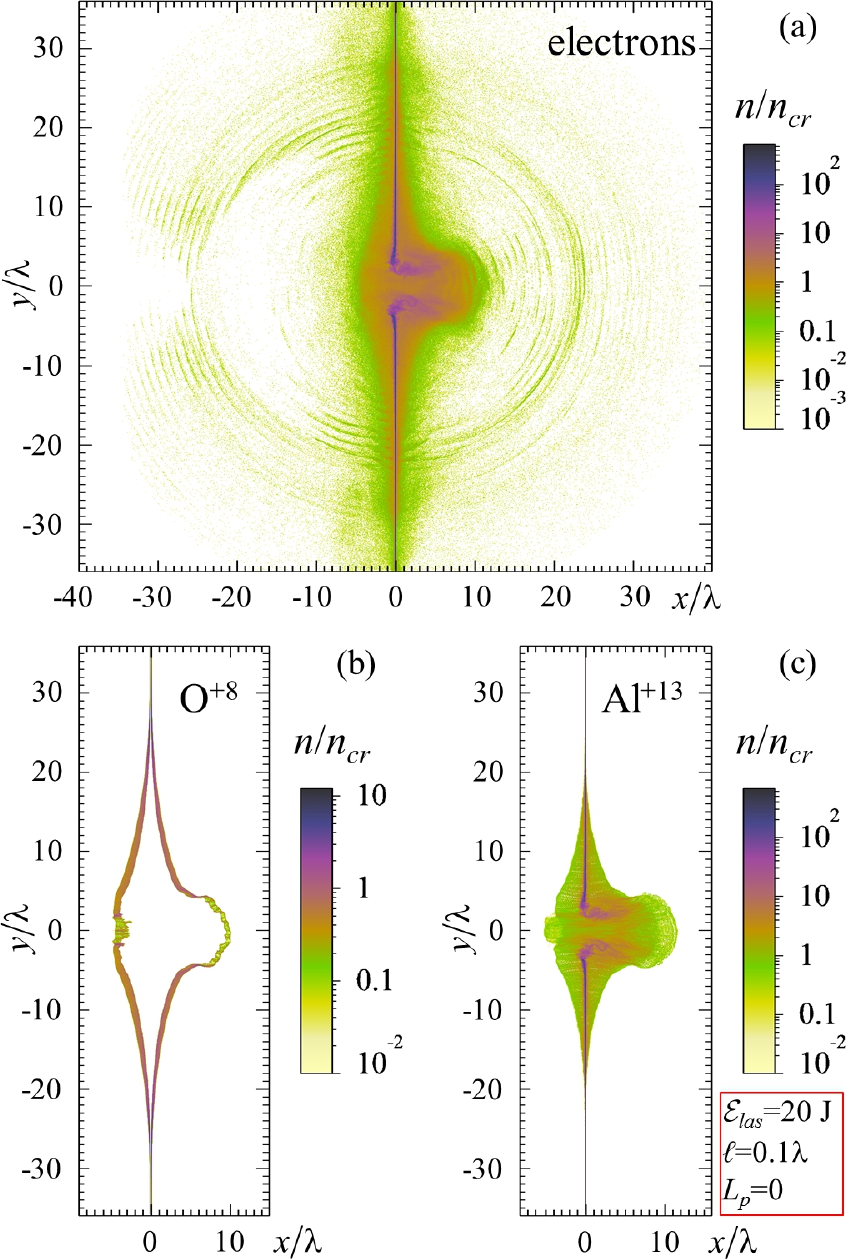}
\caption{\label{fig:220-dens}
(Color online).
Electron (a) and ions (b,c) densities
in the case without {\it preplasma} (formally $L_p=0$) for $\ell=0.1\lambda$, 
obtained with the 20 J {\it main pulse}
at $t=100 \lambda/c$.
}
\end{figure}

\begin{figure}[tbph]
\includegraphics[width=0.8\columnwidth]{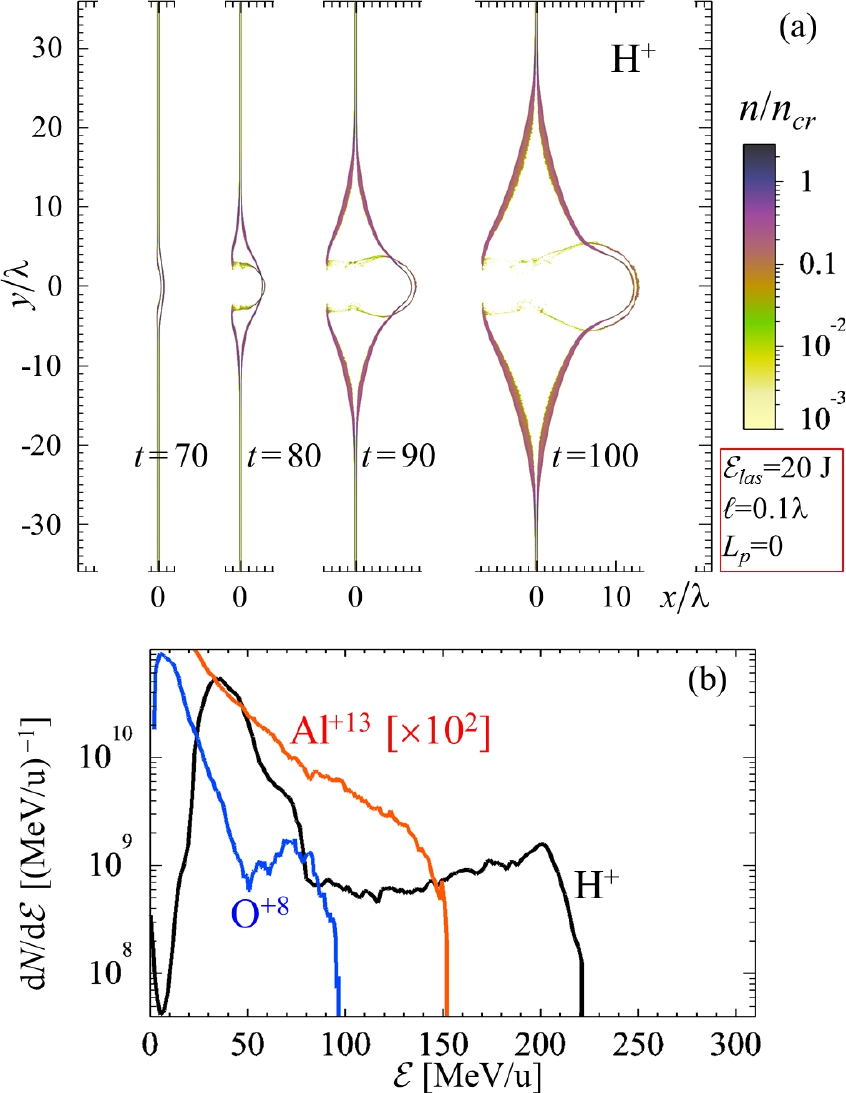}
\caption{\label{fig:220-H}
(Color online).
Proton density (a)
observed in the same case as Fig. \ref{fig:220-dens}
at $t=70,80,90,100 \lambda/c$.
(b) Energy spectra for ions at $t=140 \lambda/c$.
The distribution for Al$^{+13}$ is multiplied by $10^2$.
}
\end{figure}

\begin{figure}[tbph]
\includegraphics[width=0.8\columnwidth]{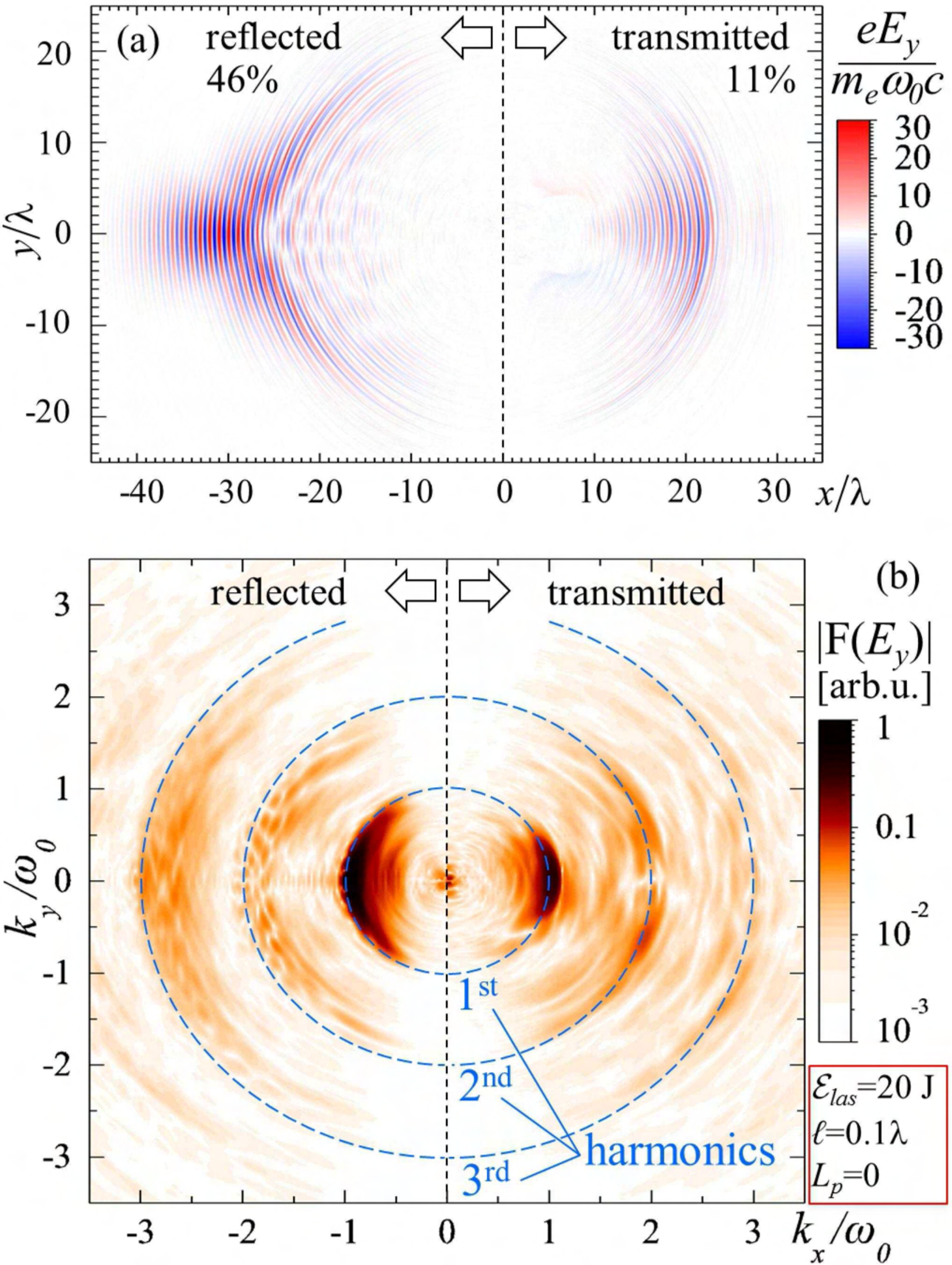}
\caption{\label{fig:220-ey}
(Color online).
Electric field $E_y$ component (a)
and the absolute value of its fast fourier transform (b)
revealing harmonics in the reflected and transmitted radiation;
observed in the same case as Fig. \ref{fig:220-dens}
at $t=100 \lambda/c$.
The reflected {\it main pulse} energy 
is 46\%, while 11\% is transmitted and the rest is absorbed.
}
\end{figure}

The results of the simulations are shown in
Figs.~\ref{fig:focus}-\ref{fig:220-ey},
where the spatial and time units are $\lambda$ and 
$\lambda/c \approx 2.67$ fs, respectively,
and
the electromagnetic field strength is shown in terms of
the dimensionless amplitude $a_0 = eE/m_e \omega_0 c$.

\subsection{Interaction scenarios}
Our simulations reveal typical scenarios 
of the {\it main pulse} evolution in the {\it preplasma}.
Propagating in the {\it preplasma}, the {\it main pulse}
loses energy creating a channel and piling up a high-density shell.
A sufficiently thick {\it preplasma} can absorb the {\it main pulse} 
almost completely well before the {\it main pulse} reaches the {\it plateau}.
In this case the ion acceleration is inefficient 
as we see in our preliminary simulations (not shown here).
For a sufficiently small {\it preplasma} scale-length,
the {\it main pulse} reaches the {\it plateau}
partially reflecting from it and partially penetrating through it.
This appears to be the necessary condition for an efficient ion acceleration
for the {\it main pulse} parameters under consideration.
In some cases a self-focusing of the {\it main pulse}
results in a significant intensification of the pulse
and its carrier frequency downshift, as seen in Fig. \ref{fig:focus}.

\subsection{Ion acceleration in targets with preplasma}
Figs.~\ref{fig:en-H}-\ref{fig:en-Al}
summarize the simulation results in terms of
the maximum ion energy per nucleon
dependence on the {\it preplasma} scale-length $L_p$ and 
the {\it plateau} thickness $\ell_p$
for different {\it main pulse} energies.
The maximum ion energy increases with the {\it main pulse} energy.
For the {\it main pulse} energy of 
${\cal E}_{las}=$ 4, 10, and 20 J
the maximum proton energy is, respectively, 129, 218, and 322 MeV;
in the case of O$^{+8}$ ions it is           42,  96, and 184 MeV/u;
and in the case of Al$^{+13}$ ions it is     33,  85, and 153 MeV/u.
The values were taken at $t=140\lambda/c$
($\approx 70$ laser cycles after the {\it main pulse}
reaches the focal plane in vacuum). 
At this time the energy growth for all cases presented
is nearly saturated (less than 3\% growth in 10 laser cycles).
For all three ion species 
under the fixed {\it main pulse} energy 
there is an optimal {\it preplasma} scale-length
which affords the highest energy.
Surprisingly it roughly corresponds to the maximum
distance between the {\it plateau} and the critical surface, $s_c$,
Fig. \ref{fig:model}(b).
For the {\it main pulse} energy of 20 J,
Fig. \ref{fig:num-H}(a) shows
the number of protons with the energy in the interval
$0.9{\cal E}_{p max} \le {\cal E}_{p} \le {\cal E}_{p max}$
(here ${\cal E}_{p max}$ is a function of $L_p$ and $\ell$,
presented in Fig. \ref{fig:en-H}(c)).
The number of protons in the 200 MeV beamlet with a 10\%
energy spread is shown in Fig. \ref{fig:num-H}(b).

Varying the {\it preplasma} scale-length $L_p$
and {\it plateau} thickness $\ell$
we observe different regimes of ion acceleration.

When the {\it main pulse} cannot penetrate a thick {\it preplasma},
fast electrons produced by the absorbed laser pulse
penetrate through the target,
creating a charge separation field at the rear of the target.
This field accelerates ions as described by
the target normal sheath acceleration (TNSA) mechanism \cite{Gurevich-P-Pitaevsky,TNSA}
or its generalization on the non-neutral sheath expansion \cite{Nishiuchi,Passoni}.
The ion acceleration occurs also from the {\it main pulse} channel
formed inside the {\it preplasma},
mostly in the transverse direction \cite{Sarkisov},
but this results in much lower ion energy than
the acceleration at the rear of the target.

When the {\it main pulse} reaches the {\it plateau},
its intensity and the {\it plateau} thickness
determine the interaction regime.
For intensity much lower than $10^{22}$ W/cm$^2$,
the {\it main pulse}
causes a Coulomb explosion \cite{TARA, DoubleLayer} of the remaining portion of the foil.
Near that threshold of intensity,
for the {\it plateau} thickness much less than $\lambda$,
the radiation pressure dominant acceleration (RPDA) comes into play \cite{RPDA},
assisted by the directed Coulomb explosion \cite{Stepan-DC}.
We note that the mechanisms providing higher ion energy, in principle,
are accompanied by other mechanisms, since
a higher threshold is reached via all lower thresholds.

In Figs.~\ref{fig:150-dens}-\ref{fig:150-ey}
and
Figs.~\ref{fig:300-dens}-\ref{fig:300-ey}
we show two cases of the 20 J {\it main pulse}
interaction with the target,
corresponding to Fig.~\ref{fig:en-H}(c).
The density, fields and frequency spectra
are shown at $t=100\lambda/c$ 
($\approx 30$ laser cycles after the {\it main pulse}
reaches the focal plane in vacuum),
in the active phase of ion acceleration.
The ion energy spectra are shown for $t=140\lambda/c$,
when the energy growth is nearly saturated.

In the first case, the {\it plateau} thickness is $\ell=0.5\lambda$
and the {\it preplasma} is relatively gently sloped, $L_p=26.5\lambda$.
The corresponding location of the critical surface is
$s_c=16.8\lambda$ for fully ionized Al$^{+13}$
or $s'_c=0$ assuming partially ionized Al$^{+3}$.
In the second case, the {\it plateau} is two times thinner, $\ell=0.25\lambda$
and the {\it preplasma} is much steeper, $L_p=10.4\lambda$;
the critical surface location is
$s_c=20\lambda$ for fully ionized Al$^{+13}$
or $s'_c=13.2\lambda$ assuming partially ionized Al$^{+3}$.
The amount of ablated material in the second case is by
$\sim 17$\% greater than in the first case.

The density distributions, 
Fig.~\ref{fig:150-dens}(a,b) and Fig.~\ref{fig:300-dens}(a,b),
reveal that
the number the electrons ejected at the distance $\ge 20\lambda$ 
from the unperturbed foil center ($x=y=0$) in the interval of $x\ge 10\lambda$
is 1.64 times greater in the second case than in the first case;
ions are accelerated mostly forward;
Al$^{+13}$ ions from the {\it skirt} acquire larger energy 
than that from the {\it plateau}.
The ion energy spectra, 
Fig.~\ref{fig:150-dens}(c) and Fig.~\ref{fig:300-dens}(c),
show that the ion energy per nucleon is roughly two times greater 
in the second case than in the first case.

\subsection{Reflection, Absorption, and Frequency Spectrum of Reflected Laser Pulse}

Fig.~\ref{fig:150-ey} and Fig.~\ref{fig:300-ey}
show that in the first case one third of the {\it main pulse} energy is reflected
in the form of several beams propagating at different angles,
while in the second case the beam is 91\% absorbed and 
8.4\% portion is reflected mainly backward.
In both cases less than 1\% is transmitted through the target.
Both the reflected and transmitted radiation is enriched
with high-order harmonics,
Fig.~\ref{fig:150-ey}(b) and Fig.~\ref{fig:300-ey}(b).
However, in the second case, producing greater ion energy,
the spectrum is much more blurred and noisy.
A small downshift of the reflected base frequency and harmonics
is noticeable in the second case,
Fig. \ref{fig:300-ey}(b).
It is $~8$\% with respect to the carrier frequency decreased to $0.9\omega_0$,
seen in Fig. \ref{fig:focus}(c).
This indicates the onset of the RPDA mechanism of 
the ion acceleration, \cite{RPDA}.
The accelerating plasma partially reflects the incident radiation
decreasing its frequency due to the Doppler effect.
The downshift corresponds to the velocity of $v=0.04c$,
and the energy of 0.77 MeV for protons
and of 21 MeV for Al ions.

\subsection{Ion acceleration in ``clean'' targets}
For the sake of comparison
we consider here the case of a target without a preplasma (formally $L_p=0$).
This corresponds to an ideally ``clean'' laser pulse
incident on an unperturbed foil.
The target consists of three layers: aluminum in the middle
and $0.01\lambda$ thick water contaminants on both sides.
We assume that all ions are fully ionized as in the analysis above.
The {\it main pulse} energy takes the values of 
${\cal E}_{las}=$ 4, 10, 15, 20 J;
the foil thickness runs through 
$l_0/\lambda = $ 0.1, 0.25, 0.5, 1.

Fig.~\ref{fig:en-H-clean} shows the maximum energy of ions
as a function of the {\it main pulse} energy, ${\cal E}_{las}$, 
and the foil thickness, $l_0$.
For the same {\it main pulse} energy
and foil thickness, $l_0=\ell$, presented in Figs.~\ref{fig:en-H}-\ref{fig:en-Al},
a ``clean'' target produces substantially less energetic ions.
The greatest value is reached for the 20 J {\it main pulse}
and $0.1\lambda$ thick foil;
for this particular case the 
density distributions, ion energy spectra, quasistatic fields,
and transverse electric field and its frequency spectrum
are shown in Figs.~\ref{fig:220-dens}-\ref{fig:220-ey}.

The {\it main pulse} partially penetrates through the target
and undergoes partial reflection, Fig.~\ref{fig:220-ey}(a),
boring a hole in the target, Fig.~\ref{fig:220-dens}(a).
The Al$^{+13}$ ions pushed by the {\it main pulse} radiation pressure
are accelerated further by a Coulomb explosion, Fig.~\ref{fig:220-dens}(c).
The protons from the rear of the foil
are also pushed by the radiation pressure and a slowly moving
electric potential of the exploding Al layer, Fig.~\ref{fig:220-H}(a).
Surprisingly,
the protons from the front side of the foil
are accelerated even more efficiently,
since they overcome the protons from the back side of the foil.
The evolving front proton layer develops a ``cocoon'' shape, characteristic for RPDA.
Fig.~\ref{fig:220-H}(b) shows the ion energy spectra.
The maximum energy of O$^{+8}$ ions appears to be less than that of Al$^{+13}$ ions.
Seen for $l_0=0.1\lambda$ and ${\cal E}_{las}\ge 10$ J
this effect is the result of 
a fast redistribution of ions during the course of the interaction.

The amount of radiation transmitted through the target
is much greater than in the case of the target with a preplasma,
Fig.~\ref{fig:220-ey}.
A well pronounced steepening of the {\it main pulse} profile
transmitted through the thin foil, Fig.~\ref{fig:220-ey}(a, right),
is the result of the relativistic transparency, \cite{Vshivkov}.
A noticeable frequency downshift by $\sim 6$\%
due to the Doppler effect
is seen in the reflected radiation, Fig.~\ref{fig:220-ey}(b, left),
while the transmitted radiation base frequency 
and harmonics are not shifted, Fig.~\ref{fig:220-ey}(b, right).
This ``red'' shift corresponds to the velocity of $v=0.031c$,
0.45 MeV for protons and 12 MeV for Al ions.

\subsection{Summary for PIC simulations}
For the efficient acceleration of ions
the {\it preplasma} must be sufficiently thin in order to allow
the {\it main pulse} to reach the remaining portion of the foil ({\it plateau}).
The most efficient acceleration in terms of the maximum ion energy
occurs when the preplasma scale-length is optimal
and the {\it plateau} is thin enough.
In this case,
due to the self-focusing of the {\it main pulse}
despite its depletion,
the radiation intensity becomes high enough
to make the plateau relativistically transparent \cite{Vshivkov}
and to render radiation pressure dominant acceleration \cite{RPDA}.

%
In order to formulate optimal conditions 
for the experimental realization of the ion acceleration
we should take into account that
in experiments it is difficult to control conditions of
the preplasma and the thickness of the remaining portion of the foil.
The process of the preplasma formation is determined
by relatively low-intensity portions of the laser pulse
which, in general, are not sufficiently stable
for ensuring reproducibility of the target condition 
at a time when the main laser pulse arrives.
Since thinner targets can be completely destroyed by the prepulse
without producing the desired ion beam,
it is safer to use foils
with initial thickness of the order of microns.
Restricting the prepulse intensity and duration 
so that the resulting preplasma is not thicker than 20 $\mu$m,
one can obtain 100 MeV protons with 4 J laser pulse,
150 MeV with 10 J pulse, and
200 MeV with 20 J pulse,
Fig.~\ref{fig:en-H}.

The maximum ion energy and the interaction scenario
correlate with the properties of the
{\it main pulse} reflection, transition and absorption.
The information on the reflected and transmitted radiation
energy and spectra can be relatively easily measured in experiments,
\cite{Pirozhkov-absorption,Streeter-2nd-order-harmonic}.
This can help to optimize the laser and target parameters during experiments.


\section{Discussion and Conclusions}
\label{sec:Conclusion}

In this paper we investigate the role of
a low-intensity prepulse accompanying the main pulse
of a petawatt class laser focused onto thin solid targets,
with respect to the laser-driven ion acceleration.
The major constituent of the prepulse is a nanosecond amplified spontaneous emission.
The prepulse heats, melts and evaporates 
a portion of an initially solid density target
creating an extended preplasma at the target front.
The preplasma scale-length is estimated using a simple analytical model
and its typical profile is found with dissipative hydrodynamic simulations.

The main pulse propagating in the preplasma
undergoes fast depletion transferring its energy to the energy of fast electrons.
A relatively thick preplasma can substantially absorb the main pulse.
Contrary to an intuitive expectation that 
such preplasma hinders the ion acceleration due to the main pulse depletion,
our analytical model predicts that in a preplasma with an optimal thickness
the femtosecond main pulse can reach the remaining portion of the target
and produce the radiation pressure dominant acceleration regime of ions.
This is in part due to the relativistic self-focusing of the main pulse 
and the resultant increase of its intensity.
Our multi-parametric PIC simulations demonstrate this possibility
and show that by optimizing the preplasma scale-length 
one can substantially enhance the laser-driven ion energy
as compared with the case of a ``clean'' laser pulse (without a prepulse)
irradiating a ``clean'' target (without a preplasma).

Having petawatt class lasers with a ``clean'' laser pulse could allow
more control, since a second lower intensity laser pulse arriving before the main
pulse could be used to create the optimum preplasma.

The maximum ion energy and other features of the laser-target interaction
correlate with the regimes of the main pulse absorption, reflection, and transition,
including high-order harmonic generation.
The analysis of the reflected and transmitted radiation energy and spectra
is an important diagnostic tool in
experimental searches for the optimal regimes of the ion acceleration.

Our results show that the ion acceleration to
a few hundred MeV per nucleon is achievable with petawatt class lasers
having a finite contrast.


\section*{Acknowledgments}
We thank  G. Korn, D. Margarone, T. M. Jeong for fruitful discussions. 
We acknowledge support of this work from MEXT and JSPS.
This work was partially supported by the joint research project 
of the Institute of Laser Engineering, Osaka University (under contract subject 2013B1-39).



\end{document}